\documentstyle[aps,eqsecnum,pre,multicol]{revtex}
\input epsf
\begin{document}
\title{\bf Effects of Confinement on Critical Adsorption:\\
Absence of Critical Depletion for Fluids  in Slit Pores.}
\author{  A. Macio\l ek$^1$  and R.Evans\\
     {\small \it H.H.Wills Physics Laboratory,
      University of  Bristol,   }\\
     {\small \it Bristol BS8 1TL, UK } \\
             N.B. Wilding\\
    {\small \it  Department  of Physics and Astronomy,
     University of Edinburgh}\\
     {\small \it Edinburgh EH9 3JZ, U.K.}}
\maketitle
\begin{abstract} 
The adsorption of  a near critical fluid confined in a slit pore is
investigated by means of  density  functional theory and by Monte Carlo
simulation for a Lennard-Jones fluid.
Our work was stimulated by recent experiments  for SF$_6$ adsorbed in a
mesoporous glass which showed the striking phenomenon
of critical depletion, i.e.  the adsorption excess $\Gamma$ first increases but 
then decreases very rapidly to negative values as the bulk critical
temperature $T_c$ is approached from above along  near-critical isochores.
By contrast our density functional and simulation results,
for a range of strongly attractive  wall-fluid potentials, show $\Gamma $
{\it monotonically increasing} and eventually saturating as
the temperature is lowered towards $T_c$ along both the critical ($\rho=\rho
_c$) and sub-critical isochores ($\rho<\rho _c)$. Such behaviour results from
the increasingly slow decay of the density profile away from the walls,
into the middle of the slit, as $T\to T_c^+$.
For $\rho <\rho _c$ we find that in the fluid the effective bulk field,
which is negative and which favours desorption, is insufficient to dominate
the effects of the surface fields which favour adsorption.
We compare this situation with earlier results for the lattice gas model 
with a constant (negative) bulk field where critical depletion was found.
Qualitatively different behaviour of the density profiles and adsorption is
found in simulations for intermediate and weakly attractive wall-fluid
potentials but in no case do we observe the critical depletion found in
experiments. We conclude that the latter cannot be accounted for by a single
pore model.
\end{abstract}

\section{Introduction}
\label{sec:intro}
The term 'critical depletion'  was introduced   in  connection with
experiments  designed to study the phenomenon  of critical adsorption 
for a pure fluid at a solid substrate~\cite{THOMMES94,THOMMES95}. 
When a  fluid
is brought to its bulk critical point in the presence 
of an attracting external wall or substrate,
for example along a critical
isochore, the amount adsorbed (adsorption) diverges as 
 $\tau\equiv (T-T_c)/T_c\to 0$. $T_c$ is the critical temperature.
Theory~\cite{FdG78} attributes  these
 divergences to the fact that  the wall causes
 a perturbation of 
 the order parameter (OP) profile $m(z) \equiv \rho(z)-\rho _c$,
  where $\rho (z)$ is the
density profile, to extend over a distance   $\sim \xi $,
the bulk correlation length, from the surface. Close to criticality,
 where $\xi \sim \mid \tau \mid ^{-\nu}$ ($\nu $ is the critical exponent), 
the OP profile
differs from its bulk value (fixed by the properties of the reservoir
far from the substrate) over  macroscopic distances and the adsorption
 can be a diverging quantity.
Fisher and de Gennes~\cite{FdG78} postulated that near criticality
 the OP profile should be described
in terms of a universal scaling function, i.e. 
 sufficiently close to $T_c$ and 
for sufficiently strongly attracting walls 
\begin{equation}
\label{eq:scalmagnet}
 m(z)= \tau^{\beta} {\cal N} \Bigl({z\over \xi}\Bigl)
\end{equation}
where ${\cal N}$ is a universal scaling function
and argued that as 
$\tau \to 0$ along the critical isochore ($\rho= \rho _c$)
 the adsorption $\Gamma$ takes on the asymptotic, universal form
\begin{equation}
\label{eq:scaladsorp}
\Gamma \equiv  \int_0^{\infty}m(z) dz \sim \tau ^{\beta}\xi \sim \tau^{\beta -\nu}
\end{equation}
where $z$ is the distance measured normal to the substrate, located at $z=0$.
 $\beta$ is the
critical exponent  describing the vanishing of the bulk OP and 
$\beta -\nu \approx -0.305$ for Ising magnets or fluids.
In   
 Ref.~\cite{THOMMES94} measurements were performed for 
SF$_6$ adsorbed on  a finely divided (colloidal)  graphite 
adsorbent (Vulcan 3${\cal G}$).
The volumetric method   employed  in these experiments 
 measured  $\Gamma $ directly on approaching the critical point along
 the critical or 
 near-critical isochores. Although $\Gamma $ increased as $\tau$ was reduced,
 consistent with~(\ref{eq:scaladsorp}), 
 close to $T_c$ ($\tau \sim 5\times 10^{-3}$)  the adsorption
 reached a maximum and decreased sharply taking on negative values 
 very close to $T_c$.
 Microgravity experiments by the same group confirmed
  these results~\cite{THOMMES94,THOMMES95}.
The critical depletion of $\Gamma$ was attributed to 
 the confining effects of  colloidal particles 
on the near critical fluid  and new experiments, designed to test this 
conjecture,  were performed
on the sorption of SF$_6$ in a mesoporous glass CPG-350, which
comprises a rigid interconnected system of mesopores with a nominal pore
diameter of 31nm.
For the rigid porous material the measured adsorption showed a very similar temperature 
dependence to that  
found for the colloidal graphite adsorbent.

These experimental results have stimulated  several 
 simulation ~\cite{SCHOEN95,SCHOEN97} and theoretical
studies~\cite{MACIOLEK98,DRZEW98} 
aimed at understanding the origin of
the striking behaviour of $\Gamma$ and, in particular,  to
answer the fundamental question as to
whether critical depletion 
is a  single pore phenomenon,  as is for example
capillary-condensation~\cite{EVANS90}
which arises from the combined effects of wall-fluid forces and finite-size 
 and which manifests itself in simple confining geometries.
The grand canonical Monte Carlo  simulations ~\cite{SCHOEN95,SCHOEN97}
of a Lennard-Jones  fluid confined between two  structureless, attractive
planar walls did indicate that the average,
 local density in the middle of  this slit-pore could fall below $\rho _c$ 
 under near-critical conditions thereby leading to depletion.
 However, it has been shown recently 
 that the depletion found in ~\cite{SCHOEN95,SCHOEN97} is an artifact
 of the simulation procedure~\cite{WILDING99}.
The theoretical studies~\cite{MACIOLEK98}  of the simplest model of a confined fluid,
 namely  the lattice gas subject to
identical surface fields located at the two walls,  revealed a 
physical  mechanism  which could cause a dramatic decrease of the adsorption
$\Gamma $  on approaching $T_c$ along near-critical isochores.
Consider the situation when the  density of the bulk reservoir is 
slightly lower than critical. 
Transcription of the lattice gas
 into the Ising model sets the chemical potential difference
\begin{equation}
\label{eq:transcr}
\Delta \mu\equiv \mu(\rho,T)-\mu(\rho _c,T)=2H,
\end{equation} 
where $\mu(\rho _c,T)$ is the chemical potential on the critical isochore
and $H$ is the bulk magnetic field.
 Thus  for $\rho<\rho _c$, $\Delta \mu=2H<0$ 
 and the  bulk
field favours the dilute (gas) phase.
If the surface fields are sufficiently attractive  that they favour 
adsorption of the dense (liquid) phase then one has a
competition between bulk and surface fields which influences the shape of the OP
profile and
hence the  behaviour of the adsorption in the slit-pore. 
When the pore is large and when  the bulk correlation length
is much smaller than the width of the pore $L$,  then
 the fluid in  the middle part of the pore
should behave as a
 bulk fluid.  For weak $H$,  the bulk OP (magnetization)
in the critical region behaves as
\begin{equation}
\label{m_b}
m_b= H\chi\sim  H \tau^{-\gamma} 
\end{equation}
 where   $\chi$ is the susceptibility.
Near the walls, on the other hand,  the fluid should behave as 
in the semi-infinite
 near critical system subject to the surface field.
 'Bulk' and 'surface' fields give diverging contributions, but
 of  opposite sign, to the adsorption $\Gamma$ for $\tau\to 0$.
 For large pore widths and for $\tau\gg \tau_0$, where
$\tau_0$ is defined by $\xi (\tau_0)\approx L$, the  adsorption can  be  approximated, for $H<0$, by 
\begin{equation}
\label{eq:Adsapprox}
\Gamma \sim {\cal A}_1\tau ^{\beta-\nu}-{\cal A}_2\mid H\mid \tau ^{-\gamma}L ,
\end{equation}
where ${\cal A}_1$ and  ${\cal A}_2$ are positive amplitudes.
Since  $\gamma >(\nu-\beta)$ 
the second term  always dominates for $T$ sufficiently close to $T_c$ 
provided $\mid H\mid$ is  sufficiently strong  and, as a consequence, 
 depletion of $\Gamma$ will occur.
 For even smaller $\tau$, such that $\tau\ll \tau_0$, the adsorption saturates 
 at a value which depends on  $H$ and on the surface field.
 The lattice gas model of a single pore predicts   critical depletion for
 $\rho<\rho_c$ only. For the case of  the critical isochore $\rho =\rho _c$,
 the bulk  field $H=0$  and as $\tau$ is reduced  $\Gamma $  first increases monotonically
 following the Fisher-de Gennes power law and eventually saturates taking
 on a positive value at $\tau=0$ $(T=T_c)$~\cite{MACIOLEK98}.
  
This  scenario was  confirmed by explicit mean-field lattice gas calculations ~\cite{MACIOLEK98}
and for two dimensional Ising
 films, with bulk  and surface fields
 of opposite sign, by  density matrix renormalization
group calculations
~\cite{DRZEW98}. In the latter case 
 it was  shown that the near-critical fluctuations can
 lead to an even richer variation of $\Gamma (\tau)$. For certain values of 
 $H$, in addition to the maximum, 
 a  minimum of $\Gamma$
appears where the correlation length approaches the pore width 
and the competition between the effect of adsorbing walls and the large susceptibility of the central region  (favouring the dilute phase) becomes particularly strong. For weak $H<0$ the results for $\Gamma (\tau)$ obtained in the lattice gas
(Ising) model of a single pore look very similar to those 
measured in the experiments  of Ref.~\cite{THOMMES94} and ~\cite{THOMMES95}. 
In the experiments for SF$_6$ adsorbed in the controlled pore glass the actual
isochores correspond to densities  lower than
critical, i.e.  $\rho/\rho _c=0.995$ and 0.999,
 so that $\Delta \mu<0$. 
 
It is tempting then to argue that since $\Delta \mu<0$ there is an effective bulk field $H<0$ which competes with the surface fields to give rise to
critical depletion. Although this is  a plausible explanation of the observed phenomenon it does not take into account the actual situation in a fluid.
For example, if the bulk density $\rho$ is fixed according  to the experimental condition of the fluid reservoir, $\Delta \mu$ as defined by 
~(\ref{eq:transcr}) will vary as $T$ approaches $T_c$. The corresponding bulk magnetic field will vary in the same way. In the present  paper we show that taking into account
the temperature dependence  of $\Delta \mu$ has a profound effect on the behaviour of confined fluids near $T_c$.
In particular we find that under  the experimental conditions of Ref.~\cite{THOMMES94,THOMMES95} a simple fluid confined in a single slit pore should not exhibit critical depletion. Rather  the adsorption should increase  monotonically as $\tau \to 0$.
This implies that an explanation of the experimental observations is still lacking.

Our paper is organized as follows.
In Sec.2 we reconsider the physical mechanism which leads to critical depletion of adsorption in the case of the lattice gas model
of a pore considered in
Ref.~\cite{MACIOLEK98} and
give a heuristic argument as to why this phenomenon should not be expected for
real fluids when the reservoir density  is fixed according to experimental
conditions. Our argument is supported by explicit calculations of $\Gamma$
using the density functional approach and by grand canonical Monte Carlo
 simulations of the Lennard-Jones fluid in a slit-pore. 
 In Sec.~\ref{sec:denfun} we report   density functional results
 obtained from  a  square gradient approximation to the free energy functional and short-ranged (contact) wall-fluid 
 potentials. Both classical and non-classical bulk free-energy densities are employed and and in the classical case
 we investigate two  forms of the free energy density, namely the  Landau 
model  free energy  and the free energy of the Lennard-Jones  fluid as obtained in Ref.~\cite{GUBBINS93}
 from an accurate fit to simulation data. 
Sec.4 describes the computer simulations  of the density profiles and adsorption
 of the Lennard-Jones fluid on the critical isochore and for a sub-critical isochore $\rho <\rho _c$. Results are presented for various strengths of the 
4-10 and 3-9 wall-fluid potentials.
We conclude in Sec.5  with a discussion 
 of the relevance of our findings for the experiments described earlier.

\section{Heuristic argument.}
\label{sec-heur}

Here we reconsider the scaling argument~\cite{MACIOLEK98}
that predicts   
critical depletion  in the lattice gas model of a single pore
 and  modify it
to incoroprate  two features that are relevant 
for the case of fluids.

First, in order to mimic the experimental situation more closely 
one should account for the fact that
  $\Delta \mu\equiv \mu(\rho,T)-\mu(\rho _c,T)$  will vary as the temperature
$T$ approaches $T_c$ at constant $\rho$.
This implies that the corresponding bulk magnetic  field $H$
should also vary with $T$ and therefore that the second term of the
approximate formula (\ref{eq:Adsapprox}) for the adsorption $\Gamma $ 
might  have a different $\tau$ dependence. This in turn 
 may affect the result
of the competition between the bulk and surface fields.

Second the lattice gas  model considered in
Ref.~\cite{MACIOLEK98} has an exact particle-hole symmetry, 
which corresponds
to the trivial symmetry under reversal  of the
field $H$ in the equivalent Ising model. 
For real fluids such symmetry is only approximate.
It is well established that the reduced symmetry of fluids
 leads to  scaling field mixing
close to the critical point~\cite{REHR73}.
To linear order in
$\tau$ and in $\mu(\rho,T) -\mu_c$, where $\mu _c\equiv  \mu(\rho_c,T_c)$,
the 
scaling fields  have the form
\begin{equation}
\label{eq:mixh}
u_H\equiv \mu(\rho,T) -\mu _c-c_1 \tau 
\end{equation}
\begin{equation}
\label{eq:mixt}
u_{\tau}\equiv \tau  +c_2 (\mu(\rho,T) -\mu_c),
\end{equation}
where the parameters $c_1$ and $c_2$ are system-dependent (non-universal)
quantities controlling the degree of field mixing.
$c_1$ is identified as the limiting critical slope of the coexistence curve,
i.e. 
$c_1/T_c=\displaystyle\lim _{T \to T_c}d\mu^{coex}(T)/dT$. 
In order  to account for the asymmetry of a real fluid near its critical point
one should identify the bulk field $2H$  with $u_H$ rather than with
$\Delta \mu$ in the scaling analysis.

The temperature dependence of the bulk field $H$  depends on  the particular
equation of state.
Consider first the the simplest possibility, i.e. the classical 
equation of state  in the
critical region given by retaining  only the leading terms of the expanded 
van der Waals (vdW) equation of state. 
 In  terms of  reduced temperature  $\tau$  and density
$r\equiv (\rho-\rho_c)/\rho _c$  the  vdW  equation of state reads
\begin{equation}
\label{eq:vdW}
\Delta \mu^*=-6r-\frac{8}{3}(1+\tau)\ln\frac{1-r/2}{1+r}+
4(1+\tau)\left[\frac{1}{1-r/2}-1\right]
\end{equation}
where  $\Delta \mu ^*=\Delta \mu/P_cv_c$, $P_c$  is the  critical pressure 
and $v_c$ is the critical volume per
molecule.
The  leading order behaviour 
  of this equation  in the near critical region  is
 \begin{equation}
 \label{eq:exvdW}
 \Delta \mu^*=6r\tau+\frac{3}{2}r^3
 \end{equation}
where we have ignored terms $O(r^4)$ and $O(\tau r^3)$ and higher.
Note that Eq.(\ref{eq:exvdW}) exhibits  particle - hole symmetry in that
 $\Delta \mu^*$ along an isotherm is 
 antisymmetric with respect to the critical isochore
 \begin{equation}
 \label{eq:symm}
  \Delta \mu ^*(-r,\tau)=-\Delta \mu ^*(r,\tau).
  \end{equation}
Moreover, for this case there is no 
 scaling field mixing and
 $u_H= \Delta \mu^*$ whose magnitude {\it decreases} linearly in $\tau$ 
as $T_c$ is approached at constant   $r$.

In order to analyse the influence of a $\tau$-dependent bulk field
on the behaviour of the adsorption $\Gamma (\tau)$ we reconsider the 
approximate  formula (\ref{eq:Adsapprox}).
For the expanded vdW equation of state  (\ref{eq:exvdW}) and
 $r<0$ ($\rho<\rho _c$) Eq.(\ref{eq:Adsapprox})
becomes
\begin{equation}
\label{eq:ApadvdW}
\Gamma \sim {\cal A}_1\tau ^{\beta-\nu}
-{\cal A}_3\mid r\mid^3 \tau ^{-\gamma}L 
-{\cal A}_4\mid r\mid L,
\end{equation}
where ${\cal A}_3=(3/4){\cal A}_2$, ${\cal A}_4=3{\cal A}_2$
and for consistency with the vdW approach the critical exponents
should take on their classical values,
$\beta=\nu=1/2$, $\gamma=1$ and the first (critical adsorption) term 
diverges  as -$\ln \tau$.
Apart from the additional temperature independent term, 
Eq.(\ref{eq:ApadvdW}) has the same form as 
 for a constant bulk field (Eq.(\ref{eq:Adsapprox})).
The additional term  does not affect 
the shape of the curve $\Gamma (\tau)$. It simply shifts  $\Gamma (\tau)$ as
a whole towards negative values 
and 
for large widths of the pore $L$  and (or) relatively large $|r|$ 
it could drive $\Gamma$  negative 
  sufficiently far from the critical point  (large $\tau$).
Closer to $T_c$ the temperature dependent terms dominate and 
the analysis of 
$\Gamma (\tau)$ as a function of $H$ and $L$ performed in Ref.~\cite{MACIOLEK98}
for constant bulk field $H$,  goes through with $|H|$ replaced by $|r|^3$.
Following Ref.~\cite{MACIOLEK98}  we  rewrite Eq.(\ref{eq:ApadvdW}) in the  form
\begin{equation}
\label{Adsapprox2}
\Gamma(\tau) = \tau^{\beta-\nu} \left[ {\cal A}_1
 - \left(\frac{\tau}{\tau_r}\right)^{\nu-\Delta}\right]
 -{\cal A}_4\mid r\mid L,
\end{equation}
where 
\begin{equation}
\label{tauH}
  \tau_r =({\cal A}_3 |r|^3 L)^{1/(\Delta - \nu)} 
\end{equation}
and we have used the exponent relation~\cite{BINDER83}
$\gamma=\Delta-\beta$ to introduce the gap exponent $\Delta$.
Once again it is understood that the exponents take their classical values.
In the region of validity of approximation (\ref{eq:Adsapprox}), i.e. for
$1\gg \tau\gg \tau_0$, with $\xi (\tau _0)\sim L$
 three different ranges of $\tau _r $
with qualitatively different behaviours of $\Gamma(\tau)$ 
can be distinguished~\cite{MACIOLEK98}:
\begin{enumerate}
\item {} $\tau_r\ll \tau_0$

In this case the first term in square brackets in 
(\ref{Adsapprox2}) dominates in
 the whole region of validity of this approximation since $\tau/\tau_r\gg 1$
 and $\nu-\Delta$ is negative.
Hence, in this region the adsorption should  increase
  monotonically as $\tau\to 0$. 
 
\item {}$\tau_r\gg 1$

This condition is equivalent to $|r|^3L\gg 1$.
In this case $\tau/\tau_r\ll 1$ throughout the critical region
 and the second term in square brackets in (\ref{Adsapprox2}) dominates.
 Hence, the adsorption is negative
 and {\it desorption} takes place despite
the presence of  adsorbing walls.

\item {}$\tau_0\ll \tau_r \ll 1$

For a given  pore width $L$ , 
the second term in square brackets in (\ref{Adsapprox2}) dominates
so long as $\tau<\tau_r$ and then $\Gamma (\tau)$ is negative. As $\tau$
 increases, $\Gamma (\tau)$ reaches  a maximum for $\tau\approx \tau _r$.
  Finally  for 
 $\tau\gg \tau_r$ the constant term in square brackets in (\ref{Adsapprox2}) 
 dominates over
the second term, and
 for such temperatures the usual Fisher - de Gennes type of adsorption
 should occur.
 
\end{enumerate}

We now consider values of parameters appropriate 
to the experiments of Ref.~\cite{THOMMES94,THOMMES95}.
Assume that $L/\sigma \sim 10^{2}$,  is of the size of the 
nominal pore diameter ( $\sigma$ is the
molecular size) of the mesoporous glass used as the adsorbent.  
For  $r=-0.001,-0.005$, corresponding
to  the two  near-critical isochores along which $\Gamma(\tau)$ was measured, 
$\tau_r \sim 10^{-7},10^{-5}$. For this value of $L$,  $\tau _0\sim 10^{-4}$
and $\tau _r \ll \tau _0$. Then according
to the  above discussion  
the adsorption $\Gamma $ should {\it increase  monotonically} 
as $T$ is lowered towards $T_c$ following these two isochores.
Even for densities that deviate more strongly  from $\rho _c$, e.g. 
 $r=-0.01$,
$\tau_r\sim 10^{-4}\sim  \tau_0$  and the condition $(3.)$
for the occurence of  depletion of $\Gamma $ might  still  not be
 satisfied. It is also very likely that for these values of $r$ and 
 $L$ the adsorption is positive for $\tau \approx \tau _0$ and hence
 for $\tau < \tau _0$ it should saturate at a positive value.
 
It is important to contrast this constant $r$ (density) scenario with the
constant $H$ lattice gas described in Ref.~\cite{MACIOLEK98}.
There depletion was observed for fields $H$ in the range -$10^{-7}$
to -$1.5\times 10^{-4}$. These were sufficiently strong to drive
$\Gamma $ negative for $\tau >\tau _0$, i.e.
while $\xi $ was smaller than $L$. $\Gamma$ then saturated at a negative
value for $\tau \le \tau _0$. In the present case, even for values of $r$ as
negative as $-0.01$, the effective field might not be strong enough
to drive $\Gamma$ negative before $\xi \sim L$ and then $\Gamma$ would saturate
at a positive value.

Our analysis so far has been based on (\ref{eq:exvdW}).
Consider now equations of state which do not incorporate the symmetry
(\ref{eq:symm}) in the $\mu - T$ plane.
For systems described by such equations of state the 'true' scaling fields
are now $u_H$ and $u_{\tau}$  and by analogy with bulk~\cite{REHR73}
 the 'true' OP $m(z)$ which satisfies the
scaling relations (\ref{eq:scalmagnet}) and (\ref{eq:scaladsorp}) is not
$\rho-\rho_c$ but rather the linear combination of the
number and entropy densities $(\rho-\rho_c)-c_2(s-s_c)$.
The entropy term in the OP does not
change the leading asymptotic behaviour of the
adsorption  for the semi-infinite system; it gives rise to a "correction term"
to  Eq.(\ref{eq:scaladsorp}) of  order
 $\tau^{1-\alpha-\nu}$, where $\alpha $ is the specific heat critical 
exponent.
In the scaling analysis of $\Gamma $ 
for the confined system, the scaling field 
 $H$ should be now replaced by $u_H$. In order to see if this can
  change the  
 behaviour of $\Gamma$ we first consider 
   equations of state that  are linear in $\tau$, as was the case in the 
 vdW equation of state. For all such equations
the 'mixed' scaling field $u_H$ 
reduces to $2H$ defined by (\ref{eq:transcr}).
This is due to the fact that the chemical potential on the critical isochore
 and the chemical potential at  coexistence have the same limiting
  slope at $T_c$, i.e. 
$c_1/T_c=\lim_{T\to T_c^{-}}(d\mu^{coex}/dT)=
\lim_{T\to T_c^{+}}(\partial\mu(\rho,T)/\partial T)_{\rho_c}$ so that
\begin{equation}
\label{eq:scH1}
u_H=\mu(\rho,T)-\mu_c-c_1\tau=\Delta\mu
+\mu(\rho_c,T)-\mu_c-c_1\tau=\Delta \mu=2H.
\end{equation}
Thus, for the classical equations of state which are linear in temperature
the reduced symmetry of the fluid does not influence the
temperature behaviour of 
the bulk field.  $2H=\Delta \mu$ differs from the leading order behaviour  only
by terms  higher in $r$. For example, in the case of the  vdW  equation of state
these are of order $r^4$ and $\tau r^3$.

If the equation of state is not linear in temperature then
\begin{equation}
\label{eq:scH2}
u_H=\Delta\mu
+\mu(\rho_c,T)-\mu _c-c_1\tau=\Delta \mu+
a_2\tau ^2+O(\tau^3),
\end{equation}
where $a_2$  is a constant coefficient.
Thus, the temperature dependence of the scaled field $u_H$, 
which now plays the role 
of the bulk field  $H$, differs from that of $\Delta \mu$, but only
by  terms higher order  in $\tau$.

We conclude that differences arising from  the
 reduced symmetry of the fluid, i.e.
'mixed' scaling fields, are  not important for the behaviour of 
the adsorption. The presence of
 higher order terms
  in  $r$ and  $\tau$  
 do not change the conclusions of 
 our analysis  performed using (\ref{eq:exvdW}). Thus our predictions
of no depletion of adsorption  
along  near-critical isochores should be
 valid for all classical equations of state.

Of course real fluids are non-classical.
Our argument can be extended using the fact that
near criticality real fluids should  obey
the scaled equation of state~\cite{WIDOM65}
\begin{equation}
\label{eq:sceqst}
\Delta \mu=r|r|^{\delta-1}D_0h(\tau/|r|^{1/\beta})
\end{equation}
where $\delta$ is the critical exponent and 
 $D_0$ is an amplitude for the power law on the critical isotherm, 
 and $h(x)$ is a scaling function. Note that $\Delta
 \mu=\mu(\rho,T)-\mu^{coex}(T)$  for $\tau<0$ and 
 $\delta=1+\gamma/\beta$. 
Although there is  no {\it a priori} theoretical  expression for $h(x)$
 the scaling function  should satisfy several conditions following from
requirements of thermodynamic stability and analyticity of the
chemical potential.
Thus,  $h(x)$ should be analytic in its range of definition 
$-1<x<\infty$, equal to 0 at $x=-1$, the coexistence curve,  and possess
an (asymptotic) series expansion  near $x=\infty$ (the critical  isochore) of the form
\begin{equation}
\label{eq:exlarx}
h(x)=\sum_{n=1}^{\infty}\eta  _nx^{\beta(\delta+1-2n)}.
\end{equation}
For small values of $x$, $h(x)$ should have an expansion of the form
\begin{equation}
\label{eq:exsmx}
h(x)=1+\sum_{n=1}^{\infty}h_nx^n.
\end{equation}
The leading temperature dependence of $\Delta \mu$
on the  near-critical isochores is given by the first term
in expansion (\ref{eq:exlarx})  and
the leading $r$ dependence is given by the first term in expansion
  (\ref{eq:exsmx}), i.e.
\begin{equation}
\label{eq:leadbeh}
\Delta \mu \sim \eta _1r\tau^{\gamma}+D_0r|r|^{\delta-1},
\end{equation}
where $D_0$ and $\eta _1$ are amplitudes.
For classical exponents this expression is consistent with  Eq.(\ref{eq:exvdW}).
Using this form for the bulk field $2H\sim \Delta \mu$ we can repeat the
 analysis performed above.
Eq.(\ref{Adsapprox2}) remains valid but now $\tau_r\sim (|r|^{\delta}L)^{1/(\Delta-\nu)}$.
 For real fluids $\delta \sim 4.78$ which means that the values of
$\tau _r$ are even smaller for a given $r$
than in the classical case. This implies that the effective bulk field
is very weak for the conditions of the experiment  and thus   depletion
of the adsorption $\Gamma $ should not occur. Rather saturation of $\Gamma$
at positive values should be expected.

\section{Results from Density Functional Theory}
\label{sec:denfun}

In this section we report the results of density functional calculations 
for the adsorption $\Gamma$
of a near-critical, simple fluid confined in a slit-like pore.
These results provide an explicit test of the  heuristic ideas given above.

Specifically, we consider a fluid  confined between two 
identical parallel adsorbing
walls located at $z=0 $ and $z=L$ and infinite 
in the $x$ and $y$ directions.
The system is in contact with a bulk reservoir  at fixed temperature $T$ and
chemical potential $\mu$.
The equilibrium profile is obtained by minimizing the
  grand potential
functional ~\cite{EVANS79}
\begin{equation}
\label{eq:gpotfun1}
\Omega[\rho]=
{\cal F}[\rho]-\int d{\bf r}(\mu-V({\bf r}))\rho({\bf r}),
\end{equation}
where  $V({\bf r})$ is the total wall-fluid  external potential 
\begin{equation}
\label{eq:expot}
V({\bf r})\equiv V(z)=U(z)+U(L-z)
\end{equation}
and $U(z)$ is the solid-fluid potential due to  a single wall. 
The equilibrium density profile  $\rho({\bf r})\equiv \rho(z)$
corresponds  to the minimum of $\Omega[\rho]$.
 We choose the simplest form for  $\Omega$
 based on the square gradient approximation
  to the intrinsic free energy functional
 ${\cal F}[\rho]$ and  model the   wall-fluid contribution 
   by a term $\Phi_s$ which depends only on the
 fluid density at contact i.e. on $\rho (0)=\rho(L)$.
In this approximation the grand-potential excess 
 per unit area is the following functional~\cite{CAHN77}:
\begin{equation}
\label{eq:gpotfun2}
\gamma[\rho]
=\frac{1}{2}\left(\displaystyle\int_0^Ldz\left[\psi(\rho)+\frac{D}{2}\left(\frac{d\rho}{dz}\right)^2\right]+\Phi_s\right).
\end{equation}
Here $\psi(\rho)\equiv \omega(\rho)+P$ is the 
excess grand-potential
 density, i.e.  $\omega(\rho)\equiv f(\rho)-\mu\rho$ is 
the grand potential density, $P$ is the  pressure and $f(\rho)$ 
is the Helmholtz
free energy density of a homogenous
fluid of density $\rho$.
 For $T<T_c$ $\psi(\rho)$ has two minima 
corresponding to the two distinct bulk
 phases. At bulk coexistence both minima are equal to zero.
The coefficient  $D$  is related to the  second moment of the
direct correlation function~\cite{EVANS79} but for simplicity we choose it to be 
density independent.
The wall-fluid term has the form
\begin{equation}
\label{eq:wallfluid}
\Phi_s=\frac{c}{2}(\rho^2(0)+\rho^2(L))
-\varepsilon _w(\rho(0)+\rho(L)).
\end{equation} 
The first term,  with $c>0$, represents a reduction of
attractive  pair interactions
between fluid particles at the surface  arising from  
exclusion of the fluid by a wall.
The second term with $\varepsilon _w >0$ measures  the strength of
 the attractive wall potential.
  Symmetry of the wall-fluid potential dictates
  that $\rho(0)=\rho(L)$ and $d\rho/dz=0$ at
 $z=L/2$.
 
It is well known that    functionals of this type 
 cannot incorporate short-ranged correlations and
 hence cannot acount for oscillations of the  density profile which occur
 near the walls~\cite{EVANS90}.  However, they should capture the main features
of  critical  adsorption  in large pores, as this phenomenon is dominated by the
 behaviour of the  profile far from the walls. Indeed they
  were succesfully employed 
 by Marini Bettolo Marconi~\cite{MARINI88} in a pioneering study of the
 effects of finite size on critical adsorption.

 Minimization of  (\ref{eq:gpotfun2}) yields the following  equation  for the density
 profile $\rho (z)$:
 \begin{equation}
 \label{eq:Eu-Lag}
 D\frac{d^2\rho (z)}{dz^2}=\frac{d\psi}{d\rho(z)}
 \end{equation}
 with boundary condition at the wall $z=0$,
\begin{equation}
\label{eq:bouncond}
D\left[\frac{d\rho(z)}{dz}\right]_{z=0}=c\rho(0)-\varepsilon _w.
\end{equation}
Eq.(\ref{eq:Eu-Lag}) has a first integral 
\begin{equation}
\label{eq:firstint}
\frac{D}{2}\left[\frac{d\rho(z)}{dz}\right]^2=\psi (\rho)+F,
\end{equation}
where $F$ is a constant of integration, independent of $z$, whose value depends
on $T$,$L$ and $\mu$. 
The function $F(L)$, which vanishes as $L\to \infty$, can be identified 
with the solvation force 
between the walls~\cite{MARINI88,EVANS85} i.e. $F(L)=-2(\partial
\gamma/\partial L)_{T,\mu}$, where $\gamma$ is the equilibrium value of
$\gamma[\rho]$. It may be determined  from the equations
\begin{equation}
\label{eq:L}
L=(2D)^{1/2}sgn(\rho(0)-\rho(L/2))\int ^{\rho(0)}_{\rho(L/2)}\frac{d\rho}{[\psi(\rho)-\psi(\rho(L/2))]^{1/2}}
\end{equation}
and
\begin{equation}
\label{eq:F}
F=-\psi(\rho(L/2),
\end{equation}
both of which follow from Eq.(\ref{eq:firstint}) along with
\begin{equation}
\label{eq:dodrow}
\psi(\rho(0))-\psi(\rho(L/2))=\frac{1}{2D}(c\rho(0)-\varepsilon _w)^2
\end{equation}
which follows from the boundary condition~(\ref{eq:bouncond}). 
The key quantity of this study, the Gibbs adsorption $\Gamma _G$  (coverage) 
is defined as
\begin{equation}
\label{eq:Gadsorp}
\Gamma _G = \int_0^L (\rho (z)- \rho _b) dz 
\end{equation}
with $\rho _b$ the density of the bulk fluid at chemical potential
$\mu$ and temperature $T$. $\Gamma _G$ satisfies the Gibbs adsorption equation
\begin{equation}
\label{eq:Gibs}
\Gamma _G=-2(\partial \gamma (L)/\partial \mu)_T,
\end{equation}
which gives, using Eq.(\ref{eq:gpotfun2}),
the following expression for $\Gamma _G$
\begin{equation}
\label{eq:exgam}
\Gamma _G= (2D)^{1/2}sgn(\rho(0)-\rho(L/2))\int^{\rho(0)}_{\rho(L/2)}\frac{d\rho(\rho(z)-\rho_b)}{[\psi(\rho)-\psi(\rho(L/2))]^{1/2}}.
\end{equation}

In order to test our predictions from Sec.2 we chose three different models 
for $\psi(\rho)$  and calculated 
 the adsorption   as a function of temperature on
 approaching  $T_c$ from above
along  {\it near-critical} isochores, i.e.  for fixed $\rho \le \rho _c$.
\begin{description}
\item{(a)} Landau model free energy.
\end{description}

In this case we expand the  grand potential density $\omega(\rho)$
and the pressure $P=-\omega(\rho _b)$
  about the critical density $\rho _c$.  In terms of reduced variables
$r=(\rho-\rho_c)/\rho_c$ and $r_b=(\rho_b-\rho_c)/\rho_c$ the dimensionless
excess  grand potential is 
\begin{equation}
\label{eq:Landform}
\psi^*(r)=\frac{a^*}{2}(r^2-r^2_b)+\frac{b^*}{4}(r^4-r_b^4)-(r-r_b)\Delta\mu^*,
\end{equation}
where $\psi^*\equiv \psi(\rho)/k_BT_c\rho_c$, 
$(k_BT_c/\rho_c)a^*\equiv a=(\partial\mu/\partial\rho)_T$ at $\rho=\rho_c$,
$(k_BT_c/\rho_c^3)b^*\equiv b=\frac{1}{6}(\partial^3\mu/\partial\rho^3)_T$
at $\rho =\rho_c$ and $\Delta\mu^*=(\mu-\mu(\rho_c,T))/k_BT_c$~\cite{NOTE}.

Such a choice for $\psi(\rho)$  corresponds to the simplest 
mean-field or Landau description of a  model fluid exhibiting particle-hole symmetry.
For the special case  $\Delta\mu=0$ (on the critical isochore $\rho _b=\rho
_c$) the integrals (\ref{eq:L})  and 
(\ref{eq:exgam}) can be performed explicitly 
in terms of 
the incomplete Jacobi elliptic
integral of the first kind~\cite{ABRAMOWITZ}. For $\Delta \mu <0$, i.e. for $\rho_b<\rho_c$, 
the relation (\ref{eq:L}) between  the order parameter at the midpoint $r_m$
and the wall separation $L$  
 can also be expressed  in terms of the elliptic
integral of the first kind (see Appendix).

For a given value of the separation $L$ between the walls  and
for fixed bulk density $\rho_b$, we  determined the reduced 
 densities $r_m$ and $r_w$ and $F(L)$    at various temperatures,
   corresponding 
  to $\tau$ between 0.1 and 0, using a graphical
construction~\cite{EVANS85} along  with 
Eq.(\ref{eq:L}). For a given $L$ there is only one solution in this range of
temperatures.
  At each temperature we  calculate numerically the integral (\ref{eq:exgam})
   for $\Gamma _G$ using the Romberg  method. 
As a check of the accuracy we  calculated $\Gamma _G$  for $r_b=0$  using  
analytical expressions for the integrals (\ref{eq:L})  and 
(\ref{eq:exgam})in terms of 
the incomplete Jacobi elliptic
integral of the first kind.

 We performed our calculations using parameters   $a^*$ and $b^*$
 obtained by fitting 
   the critical temperature  and the
 critical density of SF$_6$, i.e. $\rho _c=3.05$  nm$^{-3}$,
  $T_c=318.7$K using
  a generalized vdW equation of state (see Ref.~\cite{MARINI88} for
 details).
 We find  $a^*=2.764\tau$ and $b^*=0.113$. The value of $D$ was taken from 
 Ref.~\cite{MARINI88}, giving $D^*=1.094$.
 The wall fields $c$ and $\varepsilon _w$ were treated  as  independent
 parameters  which were varied in order to examine the influence of the strength of the 
 wall potential on the behaviour of $\Gamma _G(\tau)$.
 
 We studied systems  with different  wall separations $L$ ranging between 
25 and 200 nm. For each  $L$ and fixed wall fields $c$ and $\varepsilon _w$
 we calculated  $\Gamma _G(\tau)$
along the critical isochore and for several near-critical isochores,
 i.e. for fixed $r_b$ between 0 and $-$0.1.

The results for $L=100$ nm, $c^*\equiv c/D=0.5$ nm$^{-1}$
 and $\varepsilon^* _w\equiv
\varepsilon _w/(\rho _cD)= 1$ nm$^{-1}$ and several values of $r_b$ are shown in
Fig.1A.  
Similar behaviour of the adsorption is found  for the other values 
of $L$ we considered. Note that the effective hard-sphere diameter resulting
from fitting the SF$_6$ data is $0.433$ nm~\cite{MARINI88}.
In order to allow for a more direct comparison with 
the analysis of Sec.2 
 and with the
 results from the lattice gas (Ising) model of Ref.~\cite{MACIOLEK98} we plot in Fig. 1 the 
 quantity 
 \begin{equation}
 \label{eq:gamc}
 {\Gamma}_{c}\equiv \int _0^L dz(\rho(z)-\rho_c)=\Gamma _G +\rho _cr_bL
 \end{equation}
  rather
 than Gibbs  adsorption $\Gamma _G$.
Our choice  for the  coefficients $c^*$ and  $\varepsilon _w$ corresponds to walls
that  attract fluid rather strongly.
 On the critical isochore, i.e for $r_b=0$,
the reduced contact density $r_w\approx 0.57$ at $T=T_c$ 
and does not change very much as the temperature is increased.
Our results show no depletion; 
for all choices of the the bulk density  the curves $\Gamma _c(\tau)$ 
increase monotonically with decreasing $\tau$.
The adsorption  first increases as $T$ approaches
$T_c$  from above and then saturates sufficiently close to $T_c$.
Except for $r_b=-0.1$, the value at saturation is positive. Moreover for the 
 isochores which are the nearest to the
critical one, i.e. for $0\le r_b\le -0.01$, the values of 
 $\Gamma _c(\tau)$ are very close  for $\tau <10^{-4}$
 and the adsorption saturates  at  $\tau\approx 10^{-5}$  
 corresponding to 
$\xi \sim L$, which agrees with the result found from the Ising model 
~\cite{MACIOLEK98}.
We have examined the behaviour of $\Gamma _c(\tau)$ for several
choices of the  coefficients $c^*$ and 
$\varepsilon^*_w$  but no qualitative differences in 
behaviour  
have been found. When the strength of the wall-fluid interaction
$\varepsilon^*_w$  decreases,
the degree of adsorption becomes smaller.

\begin{description}
\item{(b)} Lennard-Jones fluid free energy.
\end{description}

As a second model  we chose  $f(\rho)$  to be  the Helmholtz free 
energy density of the Lennard-Jones (LJ) fluid as given by the empirical,
 modified Benedict-Webb-Rubin (MBWR) equation of state~\cite{GUBBINS93}:
\begin{equation}
\label{eq:MBWR}
 f^*(\rho^* )\equiv
\rho^*A^*(\rho^*)=\rho^*\left(\sum_{i=1}^{8}\frac{a_i\rho^{*i}}{i}+\sum_{i=1}^6b_iG_i\right)+\rho^*A^*_{id}\end{equation}
where $f^*(\rho^*)\equiv f(\rho)\sigma ^3/\epsilon$
and  $\rho^*= \rho\sigma ^3$.   $\epsilon $ is the LJ well depth and
$\sigma $ is the LJ atomic diameter.
$A^*_{id}$ is the ideal gas term, $a_i$ and $b_i$ are nonlinear functions
of temperature and $G_i$ are functions of the density $\rho^*$ (see Ref.~\cite{GUBBINS93}).
The MBWR  equation is a classical equation of state  obtained by fitting to 
simulation data. It  describes  the near-critical  region 
of a LJ fluid quite accurately. Since it is  non-linear in temperature 
it does not possess the particle-hole
 symmetry of (\ref{eq:symm}).

In order to calculate the adsorption in this model we proceed along the lines
described above for the  
the Landau model but now we perform all the integrals numerically.
We fit the parameters  to the critical temperature
and critical density of SF$_6$ and the parameter $D$ is obtained 
 from the relations  $a=(\partial \mu/\partial\rho)_{Tc}$ at $\rho={\rho _c}$  
 and
\begin{equation}
\label{eq:relD}
a/D=\xi _0^{-2}\tau^{2\nu},
\end{equation}
with $\nu=1$, the classical value.  $\xi_0$, the correlation-length
amplitude, is set equal to the experimental value $0.2$ nm for SF$_6$.
We find $D/(\sigma^3\epsilon)= 0.364$ nm$^2$ and $\sigma=0.467$ nm.

Fig.1B  shows the results for $\Gamma _c$ 
for the wall separation $L=50\sigma$
 and $r_b$ between 0 and $-$0.15. The wall fields are chosen to be equal to 
$c^*\equiv c/D=0.5$ nm$^{-1}$ and $\varepsilon^*_w\equiv
\varepsilon_w/(\rho _cD)=0.75$ nm$^{-1}$, which yields
 a contact density $r_w\approx 0.51$ at the critical point. 
The shapes of the curves
are very similar to those for the Landau model.
The absolute values of $\Gamma _c$ are smaller as $L$ is much smaller
than the value used for the Landau model. Note that
for this  wall separation the density of the bulk reservoir must
be further removed from $\rho _c$ in order to shift the whole adsorption curve
below zero;
for $r_b=-0.15$, $~\Gamma _c$ still saturates at a positive value.

\begin{figure}[h] 
\setlength{\epsfxsize}{8.0cm}
\centerline{\mbox{\epsffile{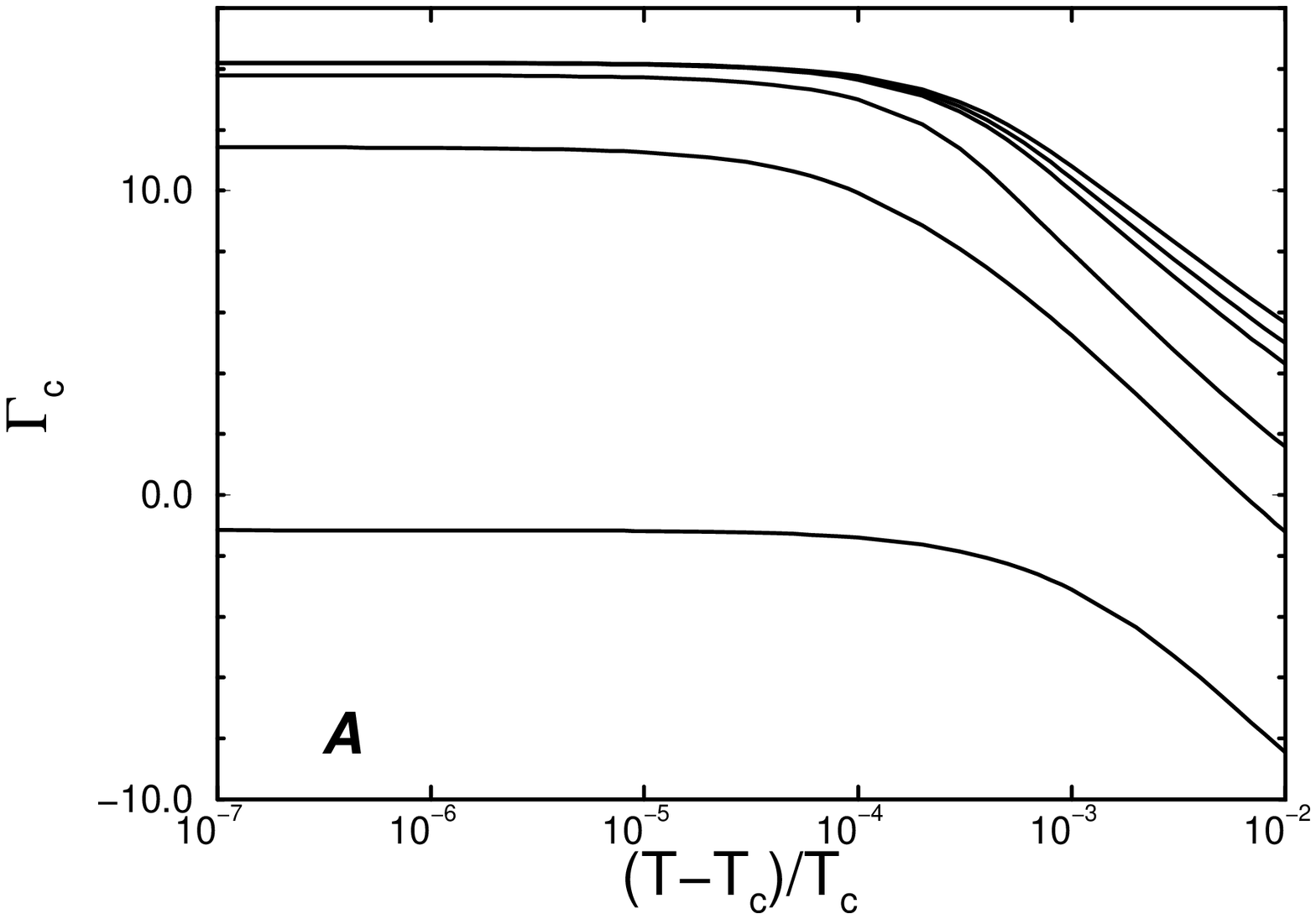}}}
\end{figure}
\begin{figure}[h] 
\setlength{\epsfxsize}{8.0cm}
\centerline{\mbox{\epsffile{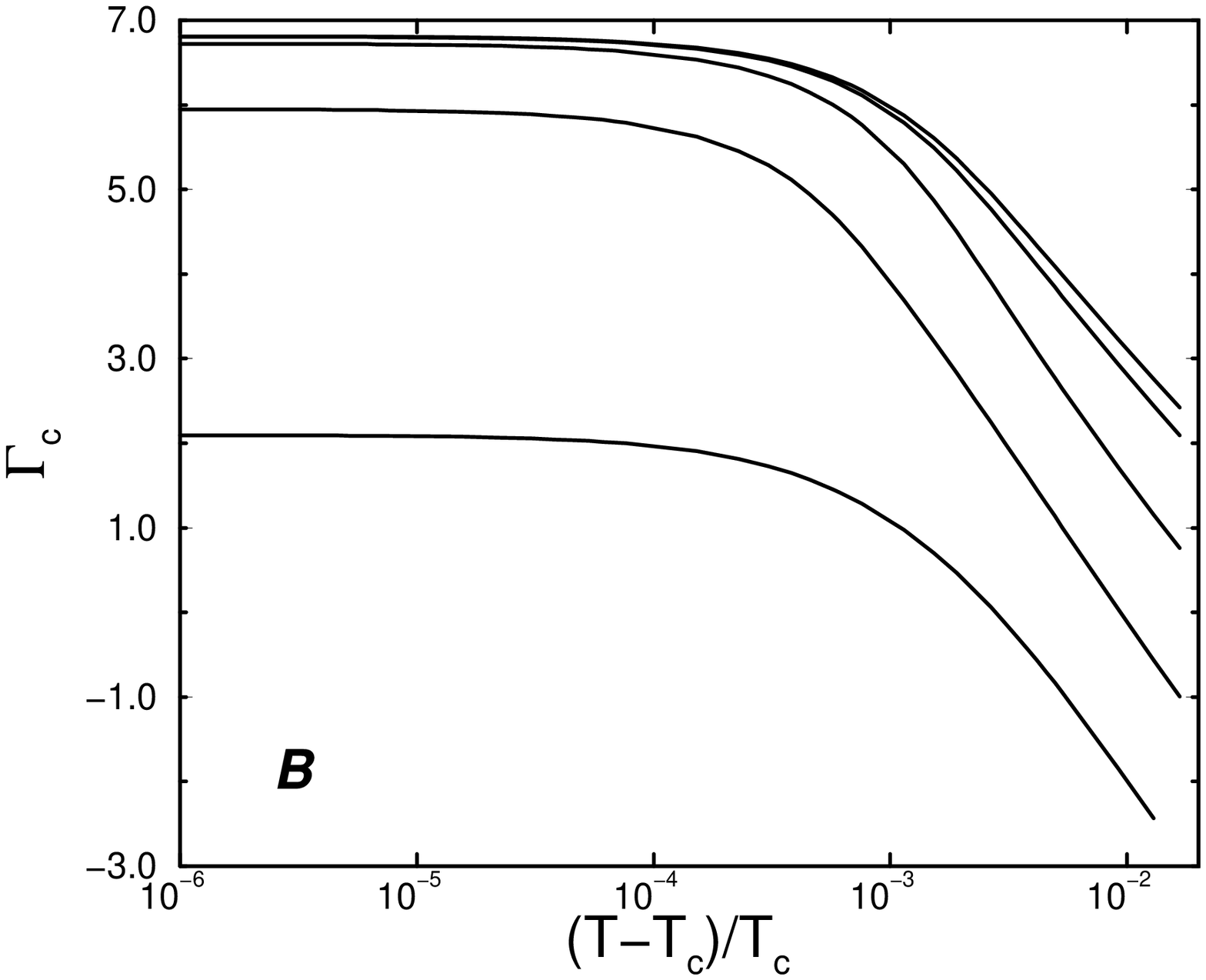}}}
\end{figure}
\begin{figure}[h] 
\setlength{\epsfxsize}{8.0cm}
\centerline{\mbox{\epsffile{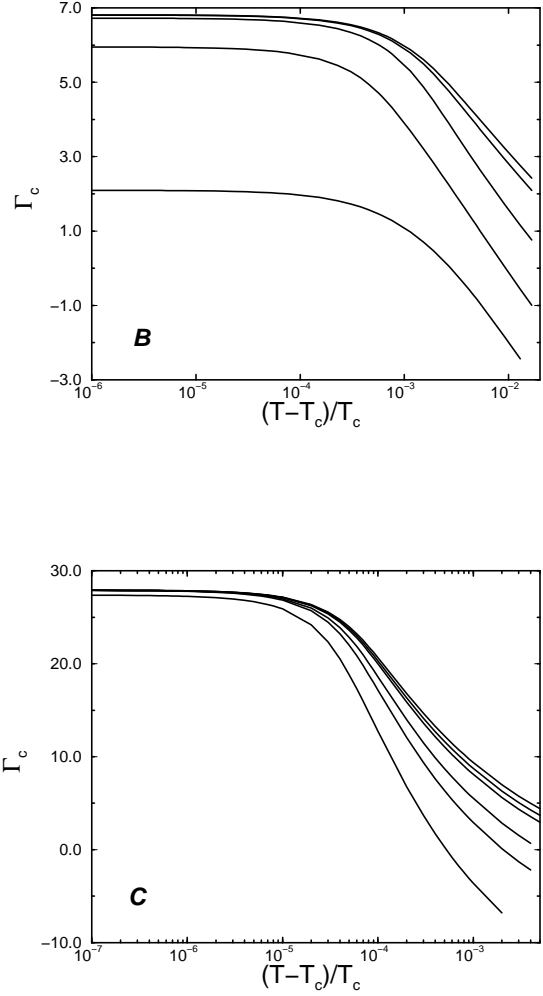}}}
\caption{Adsorption $\Gamma _c(\tau)$ (in units of nm$^{-2}$)
defined by Eq. (\ref{eq:gamc}) as
a function of the deviation from the bulk critical temperature,
 calculated in three different models (A) Landau model for
 wall separation  $L=100$ nm,  surface field $\varepsilon^* _w=1$ nm$^{-1}$
and  surface coupling $c^*=0.5$ nm$^{-1}$.
The curves correspond to different values of the reduced bulk reservoir density 
$r_b=(\rho-\rho_c)/\rho_c$. From the top  to the bottom: $r_b=0,
-0.005, -0.01,
-0.03, -0.05, -0.1$; (B)  model with the free energy density of the LJ fluid
for  wall separation  $L= 50\sigma$, where $\sigma $ is the LJ atomic
diameter, surface field $\varepsilon^* _w=0.75$ nm$^{-1}$
and  surface coupling $c^*=0.5$ nm$^{-1}$.
The curves correspond to the following  values of  
$r_b$: $r_b=0, -0.005, -0.01,
-0.1, -0.15$; (C) Fisk-Widom model for the same $L$, $\varepsilon _w^*$, $c^*$
and $r_b$ as  in the Landau model.}
\label{fig:ads}
\end{figure}

\begin{description}
\item{(c)} Fisk-Widom free energy.
\end{description}

In order to incorporate non-classical critical exponents we employ the simplest
possible  approach that goes beyond  mean field, i.e. 
the Fisk - Widom functional~\cite{FISK69}. This has  the form of (\ref{eq:gpotfun2})
with the grand potential density (\ref{eq:Landform}) replaced by
\begin{equation}
\label{eq:Wid}
\psi^*(r)=\frac{a^*}{2}(r^2-r^2_b)+\frac{b^*}{\delta+1}(r^{\delta+1}-r_b^{\delta+1})-(r-r_b)\Delta\mu^*,
\end{equation}
where $\psi^*\equiv \psi(\rho)/P_c$, 
$(P_c/\rho_c^2)a^*\equiv a=(\partial\mu/\partial\rho)_T$ at $\rho=\rho_c$,
$(P_c/\rho_c^{\delta+1})b^*\equiv b$
 and $\Delta\mu^*=(\mu-\mu(\rho_c,T))\rho _c/P_c$.
The parameter $a$ now vanishes as the
 'exact' inverse compressibility: $a\sim \tau^{\gamma}$ and the dimensionless
 quantities are defined in a slightly different way from those in the Landau
 model - see below (\ref{eq:Landform}). Following Ref.~\cite{MARINI88} we
 invoke rational approximants for the 
 critical  exponents, i.e. $\gamma=4/3,~ \nu=2/3$, $~\beta=1/3$ and $\delta=5$.
 In this approximation the specific heat exponent $\alpha$ and the
 correlation function exponent $\eta$ are equal to zero and the coefficient
 $D$  can then be treated as a constant, whereas in reality $D$ diverges
 as $\tau^{-\eta\nu}$. We define $D^*\equiv D\rho_c^2/P_c$, which has dimension
 length$^2$. 
 
 As for the Landau model, the integrals in the formulae for $L$ and $\Gamma _G$
 can be  expressed in terms of elliptic integrals of the first kind
 for the special case of the critical isochore $\Delta \mu=0$ (see Appendix).
We use these formulae  to calculate  $\Gamma _c$ along the critical 
isochore. For
$\Delta\mu\ne 0$  we  evaluate the integral for  $\Gamma$ numerically.
The parameters $D^*$, $a^*$ and $b^*$  were obtained, following Ref.~\cite{MARINI88}, from the
 Missoni-Levelt Sengers-Green nonclassical equation of state and from
relation (\ref{eq:relD}). These were fitted, as in the previous models,
 to the experimental 
values of $T_c$, $\rho _c$ and  $\xi_0$ for SF$_6$. The critical pressure
$P_c=3.7605$ MPa.

Fig.1C shows our results for $\Gamma _c$ calculated for the 
same wall separation $L=100$ nm,  the same values of the
wall fields $c^*=0.5$ nm$^{-1}$ and $\varepsilon^*_w= 1$ nm$^{-1}$ 
and the same choices of the reduced bulk densities $r_b$ as the results  obtained 
from the Landau model and shown in  Fig.1A.
Again, there is no depletion and the overall form  of the results for
the adsorption is  similar to that  for the Landau model and for the model
with the LJ fluid free energy density.
The adsorption is  stronger since $\beta-\nu=-(1/3)$ for this model
rather than 0 (logarithmic
increase) for mean field. After saturation has set in, i.e. for $\tau\le
2\times 10^{-6}$, $\Gamma _c$ is almost independent of $r_b$.
From the plot of the  midpoint density $r_m$ as a function of temperature 
(Fig.2) we see that the temperature dependence of the adsorption along the
isochores mimics the temperature dependence of $r_m$.
$r_m$ takes on a value equal to $r_b$ for large $\tau$
and then increases  as $\tau \to 0$.
This behaviour is completely  different from the case of the 
of the lattice-gas (Ising) model with constant
negative bulk field in which the magnetisation becomes more negative as $T_c$
is approached from above. 

\begin{figure}[h] 
\setlength{\epsfxsize}{8.0cm}
\centerline{\mbox{\epsffile{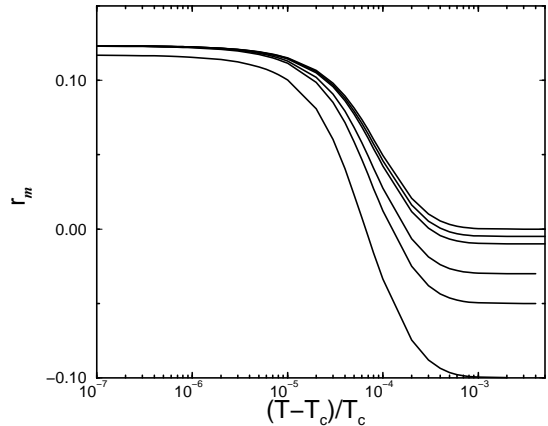}}}
\caption{The reduced midpoint density $r_m=(\rho (L/2)-\rho _c)/\rho _c$ 
 as a function of  the deviation  from
the bulk critical temperature
 calculated in the Fisk-Widom model for the same parameters as in Fig. 1C.}

\label{fig:rmid}
\end{figure}

\section{Simulation studies of a Lennard-Jones fluid}

\subsection{Computational details}
\label{sec:comp}

We have performed Monte Carlo simulations of a simple one
component fluid, interacting via an interparticle potential of the
Lennard-Jones (LJ) form:

\begin{equation} 
U_{LJ}(r)=4\epsilon\left[\left (\frac{\sigma}{r}\right)^{12}-\left(\frac{\sigma}{r}\right)^{6}\right ]\;. 
\label{eq:lj} 
\end{equation} 
Here, as in Sec.~\ref{sec:denfun}, $\epsilon$ measures the well depth of the potential, while 
$\sigma$ sets the length scale.  As is customary in
simulations of systems whose interparticle potential decays rapidly
with particle separation, the LJ potential was truncated in
order to reduce the computational effort. In accordance with many
previous studies of the LJ system, the cutoff radius was chosen to be
$r_c=2.5\sigma$, and the potential was left unshifted. No corrections
were applied to account for the effects of the truncation.

The simulations were performed within the grand canonical ensemble
\cite{ALLEN87,FRENKEL96}, permitting fluctuations in the total particle
number $N$. Two distinct geometries were studied:

\begin{enumerate}

\item[{\bf(A)}] A fully periodic cubic system of volume $V=L^d$.

\item[{\bf(B)}] A slit-pore geometry in which the fluid is confined to
a cuboidal simulation cell of dimensions $L_x\times L_y\times L$ (with
$L_x=L_y$) having structureless hard walls in the planes $z=0$ and
$z=L$, and periodic boundary conditions at the cell boundaries in the
$x$ and $y$ directions parallel to the walls.

\end{enumerate}

Consider the behaviour of the configuration  averaged local number density
$\rho({\bf r})$ in these systems. In geometry (A), translational
invariance ensures that $\rho({\bf r})$ is independent of the position
vector ${\bf r}(x,y,z)$, and one has simply $\rho({\bf r})=\rho$, i.e. the
configuration averaged number density. In the absence of finite-size effects,
$\rho$ is completely determined by the imposed values of the chemical
potential $\mu$ and the temperature $T$. By contrast, in geometry (B)
the presence of the walls at $z=0$ and $z=L$ break the translational
symmetry in the $z$ direction giving rise to a one-dimensional density
{\em profile} $\rho({\bf r})=\rho(z)$
representing the configuration averaged local number density at a given $z$.
The precise form of this profile depends not only on $\mu$ and $T$, but
also on the details of the fluid-wall interaction. In this work, we
assume that fluid particles interact with a single wall via a long ranged
potential having one of either two forms:

\begin{mathletters}
\label{eq:wpots}
\begin{equation}
U_{4-10}(z)=4\epsilon f_{4-10}\left[ \frac{2}{5}\left(\frac{\sigma}{z}\right)^{10}-\left(\frac{\sigma}{z}\right)^{4}\right ]\;\;,
\end{equation}

\begin{equation}
U_{3-9}(z)=4\epsilon f_{3-9}\left[ \frac{2}{15}\left(\frac{\sigma}{z}\right)^{9}-\left(\frac{\sigma}{z}\right)^{3}\right ]\;,
\end{equation}
\end{mathletters}
where $f$ is a parameter that tunes the strength of the fluid-wall
interactions relative to those of the fluid interparticle interactions.
The total wall-fluid potential is then given by (\ref{eq:expot}).
We note that $U_{4-10}(z)$ models a wall which is assumed to comprise a
single plane of LJ particles, while $U_{3-9}$ models a wall
that fills the half space \cite{ISRAEL}. Since both of these wall
potentials decay considerably less rapidly with increasing separation
than the LJ interparticle potential of eq~\ref{eq:lj}, no potential
truncation was applied.

\begin{figure}[h] 
\setlength{\epsfxsize}{8.0cm}
\centerline{\mbox{\epsffile{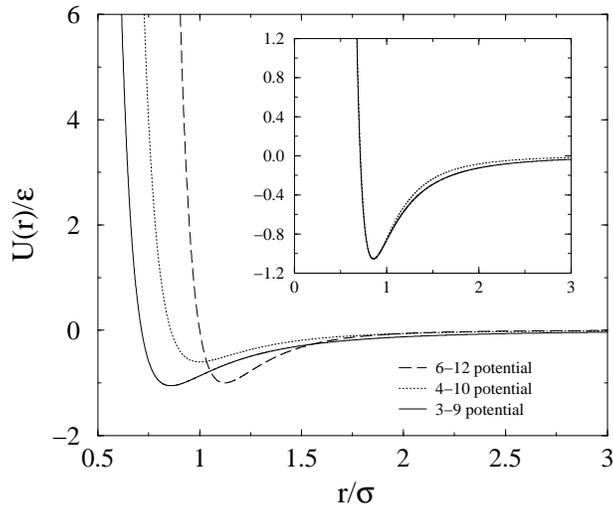}}}

\caption{Comparison of the potentials of eqs.~\protect\ref{eq:lj} and
\protect\ref{eq:wpots} for the case $4\epsilon=f=1$. The inset shows the
result of scaling the well depth of the $4$-$10$ potential to equal that
of the $3$-$9$ potential, and translating along the abscissa until the
minima coincide.}

\label{fig:pot}
\end{figure}

It is instructive to compare the forms of the two types of fluid-wall
potentials (eq.~\ref{eq:wpots}) both with one another, and with the
Lennard-Jones ($6$-$12$) interparticle potential (eq.~\ref{eq:lj}).
This comparison is made in fig.~\ref{fig:pot} for the case $4\epsilon =f=1$. The
relative range of the two wall potentials is exposed by the inset which
shows the result of first scaling the well depth of the $4$-$10$
potential to equal that of the $3$-$9$ potential, and then translating
along the abscissa until both minima coincide. In this representation,
one sees that the $3$-$9$ potential is both longer ranged and exhibits a
broader minimum than the $4$-$10$ potential---a fact which will aid the
interpretation of the simulation results presented below.

\subsection{Determining the critical isochore}
\label{sec:crit_iso}

As far as practicable, our simulation strategy has been to try to mimic
the experimental adsorption studies of ref~\cite{THOMMES94} in which
$\Gamma(\tau)$ was measured along the bulk critical isochore as $T_c$
was approached from above. A prerequisite in this regard is an accurate
knowledge of the locus of the bulk critical isochore. Within our grand
canonical simulation framework, this is specified by the function
$\mu(\rho_c,T)$. For the LJ fluid (with $r_c=2.5\sigma$), $\rho_c$ is
known accurately from a previous finite-size scaling (FSS) study, as
are estimates for $\mu_c$ and $T_c$  \cite{WILDING95_1}. Hitherto,
however, no accurate estimates for the critical isochore itself have
been reported. Accordingly a new set of simulations were performed to
determine $\mu(\rho_c,T)$ for a range of super-critical temperatures.
These simulations were carried out using simulation geometry (A)
described above.

The procedure adopted for estimating $\mu(\rho_c,T)$ is detailed below
and involves determining, for a given temperature $T$, that value of the
chemical potential for which the measured density matches the known
critical point value. Unfortunately this task is complicated by
finite-size effects. Well away from the critical point the correlation
length $\xi$ is small, and provided the linear dimension of the periodic
simulation box $L\gg\xi$, the finite-size function $\mu_L(\rho_c,T)$
will provide a reliable estimate for the bulk isochore $\mu(\rho_c,T)$.
As the critical point is approached, however, the correlation length
grows until it is comparable to, or greater than the system size $L$. In
this regime, one expects that estimates of $\mu_L(\rho_c,T)$ will
deviate systematically from the limiting bulk form.

In principle, finite-size scaling (FSS) methods can be employed to
obtain estimates for bulk quantities from simulations of finite size
\cite{PRIVMAN90}. Unfortunately their application {\em near} the
critical point is rather less straightforward than {\em at} the
critical point. The difficulties stem from crossover effects associated
with non-zero values of the two relevant scaling fields \cite{REHR73},
$u_H$ and $u_\tau$, which control deviations of the number density and
energy density from their critical point values-see Eqs.~(\ref{eq:mixh}) 
and (\ref{eq:mixt}).
 Small but
finite values of these fields result in a large but finite correlation
length $\xi$ which, owing to computational restrictions on the range of
accessible system sizes, renders the limit $L\gg\xi$ (in practical
terms) unattainable. One is therefore forced to attempt to extrapolate
to the thermodynamic limit using data from system sizes for which
$L\lesssim\xi$. However, such an extrapolation is fraught with
complications since it requires prior knowledge of the universal
scaling functions (and associated non-universal amplitudes) controlling
the crossover to the thermodynamic limit as both $|u_\tau|$ and $|u_H|$
are increased. To our knowledge, accurate forms for these scaling
functions are not available.

In view of these difficulties, we have not attempted a full FSS
analysis of the critical isochore. Instead we have simply determined 
$\mu_L(\rho_c,T)$ for the largest accessible system size and used this
as our estimate for the bulk function $\mu(\rho_c,T)$. Notwithstanding
the lack of a FSS analysis, there are formal grounds for believing that
for the special case of the critical isochore, finite-size effects are
smaller than on any other near critical isochore. To see this one must
consider the effect on observables of finite values of the scaling
field $|u_H|$. Specifically, we focus on the  ordering
operator conjugate to $u_H$, which is given by ${\cal M}\simeq
(\rho-c_2s)$, representing some particular linear combination of the
number and entropy densities \cite{WILDING92}. 

The dependence of $\delta{\cal M}={\cal M}-{\cal M}_c$ on $|u_H|$
differs between the thermodynamic and FSS limits. For the former case
one has $\delta{\cal M} \sim u_H^{1/\delta}$, while for the latter,
$\delta{\cal M}_L\sim u_HL^{\gamma/\nu}$. It follows that for the special case  $u_H=0$,
estimates of $\delta{\cal M}_L$ will agree with the bulk value of
$\delta{\cal M}$ for all $L$, i.e. exhibit no finite-size dependence.
For non-zero  
$|u_H|$, however, there will be a finite-size error
$\delta{\cal M}_L-\delta{\cal M}$. Specifically,  for a {\em given $L$}
one expects  that as $|u_H|$ is increased from zero, $\delta{\cal M}_L-\delta{\cal M}$ initially increases  like $|u_H|^{1/\delta}$, but slows
with increasing $|u_H|$, reaching a maximum at some $|u_H|\sim L^{-\beta\delta/\nu}$. Thereafter, further increase in $|u_H|$ lead to a decrease in the magnitude of $\delta{\cal M}_L-\delta{\cal M}$ until, for $\xi\ll L$, measurements of $\delta{\cal  M}_L$ again agree with $\delta{\cal M}$. Thus as $|u_H|$  is increased from zero, the magnitude of the
finite-size error $\delta{\cal M}_L-\delta{\cal M}$ associated with a
given choice of $L$, first increases from zero, reaches a maximum and then
falls back to zero.

As far as the number density itself is concerned, one finds
\cite{WILDING95_1} that on the line $u_H=0$,  $\rho$ converges rapidly
to its limiting value with increasing $L$ like $\rho_L-\rho\sim
L^{-(1-\alpha)/\nu}$. For nonzero $|u_H|$, the dominant source of
finite-size error is that described above. Thus for a given $L$,  the
finite-size error in the measured number density is minimised for state
points on the line $u_H=0$, representing the analytical
continuation of the coexistence curve to $T>T_c$. Given that the
critical isochore and the line $u_H=0$ meet at the critical point and
separate from one another only weakly as $T$ increases from $T_c$ (a
feature that we have confirmed numerically), one can expect that the
scaling field $u_H$ is generally very small on the critical isochore
(see also Sec.2) and hence that the system size dependence of $\mu_L(\rho_c,L)$ is less
than it would be on any sub- or super-critical isochore within the
critical region. 

\begin{figure}[h] 
\setlength{\epsfxsize}{8.0cm}
\centerline{\mbox{\epsffile{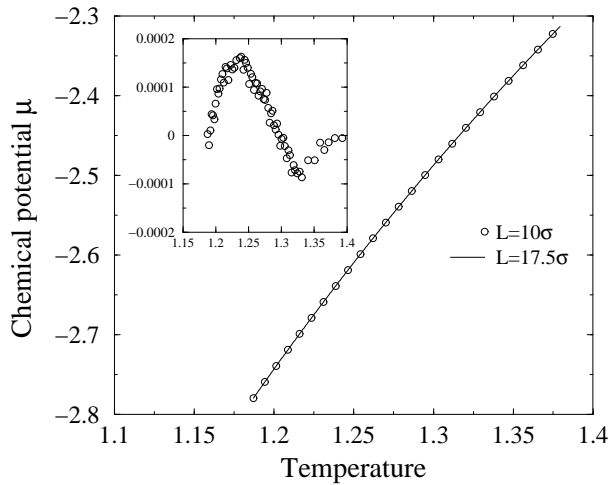}}}

\caption{The measured isochore $\mu_L(\rho_c,T)$ for the two periodic system
sizes $L=17.5\sigma$ and $L=10\sigma$. The inset displays the difference
between the two estimates. $\mu $ is the absolute chemical potential 
(in units of $k_BT$)  subject to the convention of choosing the thermal 
wavelength $\lambda=1$ in the general definition (see Ref.~\protect\cite{ALLEN87}).
Results are given in terms of LJ reduced units.}

\label{fig:isochore}
\end{figure}

To determine the finite-size critical isochore, the following procedure
was employed. Taking a large, cubic, periodic simulation cell of linear size
$L=17.5\sigma$, runs were performed at the bulk critical point, for
which \cite{WILDING95_1} in reduced  LJ units $T_c=1.1876(3), \mu_c/k_BT_c=-2.778(2),
\rho_c=0.3197(4)$. At this density, the system contains, on average,
$N=1715$ particles.  Histogram reweighting \cite{FERRENBERG88} was then
employed to estimate for $T=1.0028T_c$, the $\mu$ value for which
$\rho=\rho_c$. A second simulation was then performed at this new state
point, followed by a further extrapolation of the results to obtain
$\mu_L(\rho_c,T)$ for the still higher temperature of $T=1.015T_c$.
This procedure was then iterated a total of six times, giving isochoric
data at temperatures $T=T_c, 1.0028T_c, 1.015T_c, 1.053T_c, 1.087T_c,
1.123T_c$. Combining all six sets of simulation data via the multiple
histogram reweighting scheme \cite{FERRENBERG88}, it was then possible
to map the entire isochore in the range $T_c<T<1.123T_c$.
Fig.~\ref{fig:isochore} shows the results, together with the estimate
of the critical isochore for a smaller system of linear size
$L=10\sigma$, determined using an identical procedure. The differences
between these two estimates of $\mu_L(\rho_c,T)$ (inset of fig.~\ref{fig:isochore}) are
less than one part in $10^4$, confirming the expectation that finite-size
effects are small. 

\subsection{Studies of the slit-pore geometry on the critical isochore}
\label{sec:pore}
Simulations of the LJ fluid confined to a mesoscopic slit pore (cf.
section~\ref{sec:comp}) have been carried out for state points along
the critical isochore of fig.~\ref{fig:isochore}. Most of our studies
are for a system of dimensions $L_x=L_y=15\sigma, L=20\sigma$, although some
results have also been obtained for a system of size $L_x=65\sigma,
L=20\sigma$ in order to gauge the magnitude of finite-size effects
associated with the wall area $A=L_x^2$. We note that  in the
experiments of ref. \protect\cite{THOMMES95} for SF$_6$ in CPG glass, the
pore diameter is about $31$ nm, corresponding to about $100$
molecular diameters.

Simulation runs comprised $10^5$ Monte Carlo sweeps for equilibration
followed by $5\times 10^6$ sweeps for data collection. Measurements of
the density profile were accumulated every $100$ sweeps with each sweep
involving $(L_x/r_c)^3$ attempted particle transfers and $(L_x/r_c)^3$
particle translations. Results have been obtained for both of the wall
potentials given in equation~(\ref{eq:wpots}), and for a range of values of
the wall potential strength $f$.

\subsubsection{The critical point}

\begin{figure}[h] 
\setlength{\epsfxsize}{8.0cm}
\centerline{\mbox{\epsffile{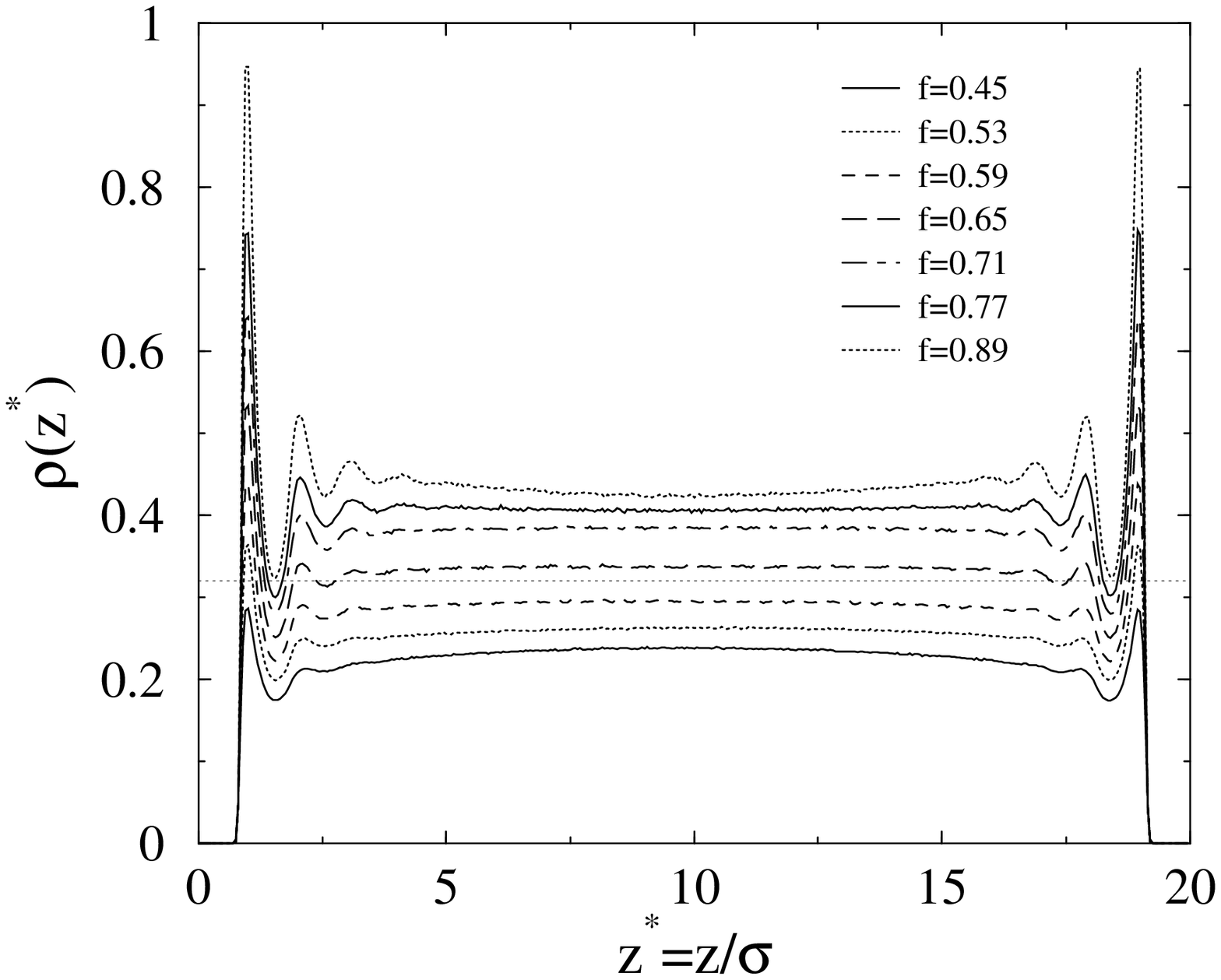}}}
\setlength{\epsfxsize}{8.0cm}
\centerline{\mbox{\epsffile{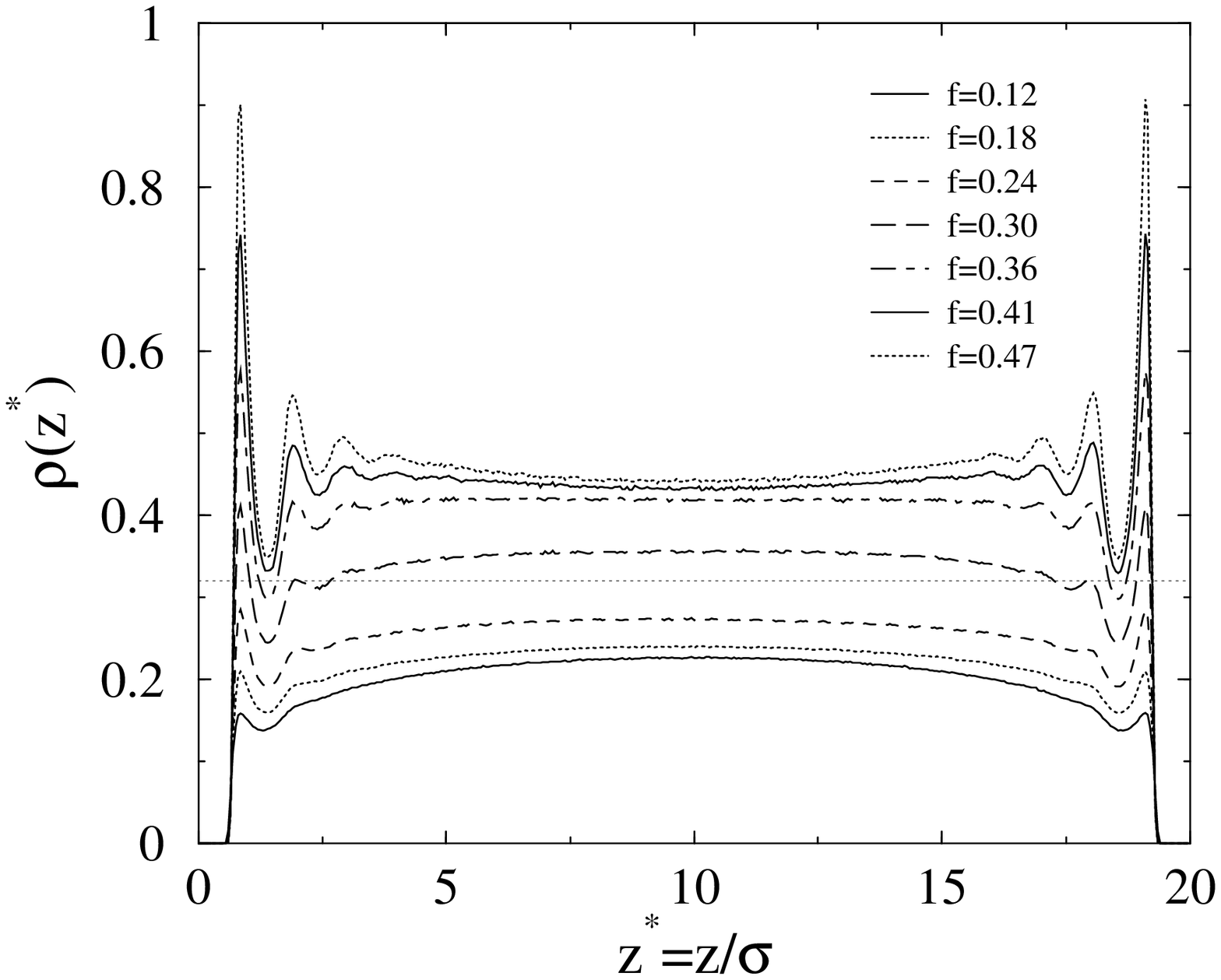}}}

\caption{{\bf (a)} The measured density profiles $\rho(z)$ (in units of
$\sigma ^3$) for the
$4$-$10$ potential at the estimated values of the bulk critical
parameters $\mu_c,T_c$. The system size is $L_x=15\sigma,L=20\sigma$.
Results for a number of wall strengths $f$ are displayed (cf.
eq~(\ref{eq:wpots}))  {\bf (b)} Same as (a) but for the $3$-$9$ potential.
In both cases the horizontal dotted line denotes the the critical
density $\rho_c$.}

\label{fig:critprofs}
\end{figure}

Density profiles $\rho(z)$ at the bulk critical parameters $\mu_c,T_c$
are shown in fig.~\ref{fig:critprofs}(a) for the $4$-$10$ potential and
fig.~\ref{fig:critprofs}(b) for the $3$-$9$ potential. In both cases
the bulk critical density $\rho_c\sigma ^3=0.3197$ is denoted by a horizontal
dotted line.

It is instructive to compare and contrast the density profiles for both
forms of wall potential. In all instances there is oscillatory
structure close to the walls arising from excluded volume `packing'
effects. The number density at the wall (as measured e.g. by the height
of the first peak) is principally controlled by the strength of the
wall potential, i.e. by the value of $f$. Larger values of this
parameter lead to a larger wall density and greater amplitude of
oscillations in $\rho(z)$. Further away from the walls, the packing
effects gradually die out and $\rho(z)$ varies smoothly with $z$.
Within this smooth region, three main regimes of behaviour (common to
both wall potentials) can be identified as $f$ is varied. We address
them in turn. 

For strongly attractive wall potentials ($f_{3-9}\geq
0.41$,$f_{4-10}\geq 0.77$), the density at the wall greatly exceeds the
critical density and the profile is {\em convex downwards} with respect
to the critical density. Thus the local density $\rho(z)$  everywhere
exceeds $\rho_c$ and decreases with increasing distance from the wall.
At no point does it fall to $\rho_c$. This latter feature is in stark
contrast to the simulation results of Schoen {\em et al}
\cite{SCHOEN95,SCHOEN97} who investigated a very similar model to that
described here. For certain fluid-wall interaction strengths,  they
reported a density profile that greatly exceeds $\rho_c$  near the
walls, but dips {\em below} $\rho_c$ in the slit middle. Recently,
however, this depletion feature has been demonstrated  to be an
artifact \cite{WILDING99}, arising jointly from systematic errors in the
simulation procedure used, and an incorrect designation of the critical
point parameters \cite{NOTE2}.

For weakly attractive fluid-wall interactions ($f_{3-9}\leq
0.24$,$f_{4-10}\leq 0.53$) the density at the wall is less than the
critical density and the profile is {\em convex upwards} with respect
to the critical density. Thus the local density $\rho(z)$ is everywhere
less than $\rho_c$ and increases with the distance from the wall. At no
point does it attain the critical density.  Tests performed at
subcritical temperatures indicate that for $f$ values in this regime,
the walls prefer the gas phase at coexistence. This is in contrast to
the systems studied experimentally in ref. \cite{THOMMES94,THOMMES95},
where the liquid wets the walls for $T<T_c$ and which, presumably
therefore, correspond (in the language of the present model) to the
large $f$ regime.

Turning now to intermediate wall strengths, the observed behaviour
is somewhat subtle. Close to the wall, the packing-induced density
oscillations span the critical density. Further away from the wall, the
magnitude of the profile curvature is generally less than in the large
or small $f$ limits. Interestingly, there exist $f$ values in this
regime for which $\rho(z)$ exceeds $\rho_c$, but the profile is {\em
concave upwards} with respect to the critical density.  Thus $\rho(z)$
exceeds $\rho_c$ and {\em increases} with increasing distance from the
wall. We shall return to discuss this unexpected finding in
section~\ref{sec:concs}.

Although both types of wall potential exhibit the same qualitative
behaviour in the three regimes of $f$ described above, differences  are
present in the detail. This is particularly true for small values of
$f$ as evidenced, for example, by a comparison of the profiles for
$f_{3-9}=0.18$ and $f_{4-10}=0.53$. For these profiles the local
densities in the slit middle are almost equal, but the wall density for
the $4$-$10$ potential is considerably greater than that for the
$3$-$9$ potential. This difference reflects the relative range of the
two potentials (as discussed in section~\ref{sec:comp}) and in
particular, the fact that to obtain a given magnitude of wall potential
at the slit middle ($z=10$), one requires $f_{4-10}=10f_{3-9}$. 

\begin{figure}[h] 
\setlength{\epsfxsize}{8.0cm}
\centerline{\mbox{\epsffile{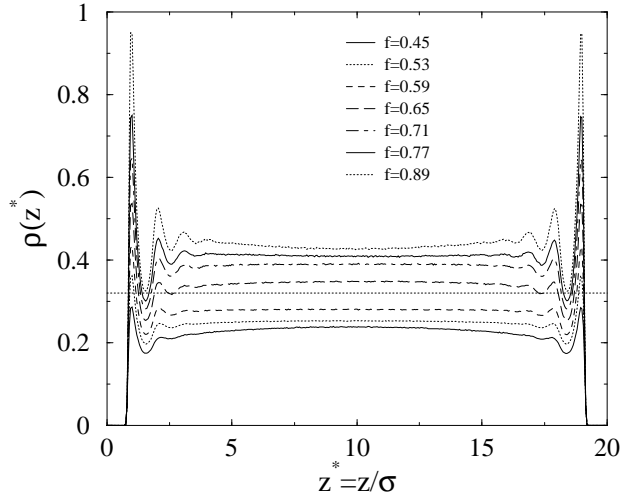}}}

\caption{As figure~\protect\ref{fig:critprofs}(a), but for a system
having $L_x=65\sigma, L=20\sigma$.}

\label{fig:wall_area}
\end{figure}

We round off this subsection with some remarks concerning finite-size
effects associated with the finite wall area $A=L_x^2$. In a simulation,
the infinite slit-pore limit $L_x/L\to \infty$ cannot be realised for all
$L$ values of interest because of bounds on the computationally
accessible system sizes.  Well away from criticality, this should
elicit no grave concern because periodic boundary conditions in
directions parallel to the walls provide a good approximation to the
thermodynamic limit. A critical system, on the other hand is always
`aware' of its boundary conditions,  by virtue of its infinite
correlation length.  Changes in $L_x$ will therefore alter the effective
range of correlations parallel to the walls, which might be expected to
couple to those perpendicular to the walls in such as way as to affect
the density profile $\rho(z)$. To investigate this possibility, we have
performed simulations in which we increased $L_x$ from the value
$L_x=15\sigma$ considered hitherto, to $L_x=65\sigma$. Owing to the high
computational cost associated with such a large simulation cell, it was
feasible to perform this comparison for only one type of wall
potential, and we have chosen the $4$-$10$ form. The results are shown
in fig.~\ref{fig:wall_area}. Comparison with those of
fig.~\ref{fig:critprofs}(a) reveal that the increase in $L_x$ engenders
only small changes in the form of the density profiles, the effect
(such as there is) being greatest for intermediate wall strengths. On
this basis it seems unlikely that the general scenario set out above
would differ in the infinite slit limit.  Nevertheless we feel that the
role of wall area on confined critical systems is an issue that
certainly merits a more systematic future investigation.

\subsubsection{Super-critical temperatures}

In this subsection we present results for the temperature dependence of
the density profile on the critical isochore. Both the $3$-$9$ and the
$4$-$10$ wall potentials have been studied in this regard. However,
since it transpires that the qualitative form of the results are
similar in both cases, we describe only those results pertaining to the
$4$-$10$ potential.

\begin{figure}[h]  
\setlength{\epsfxsize}{8.0cm}
\centerline{\mbox{\epsffile{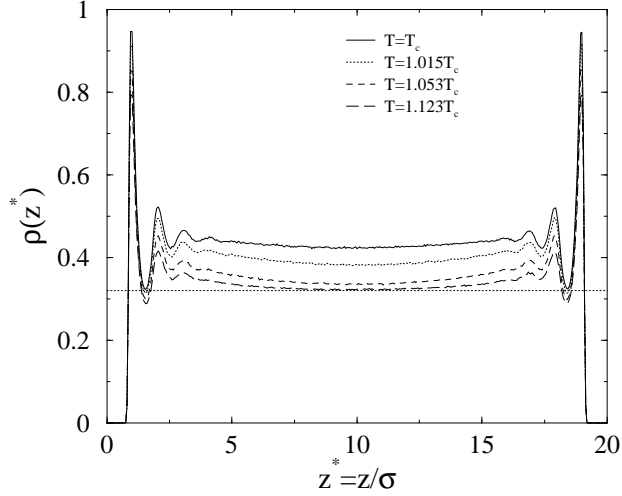}}}

\caption{The measured density profiles $\rho(z)$ (in units of $\sigma ^3$)for the $4$-$10$
potential with $f_{4-10}=0.89$ at a selection of temperatures along the
critical isochore of figure\protect\ref{fig:isochore}. The
critical density $\rho_c$ is denoted by a horizontal dotted line.}

\label{fig:super_crit}
\end{figure}

Fig.~\ref{fig:super_crit} shows the forms of $\rho(z)$ at a selection
of temperatures along the critical isochore of
fig.~\protect\ref{fig:isochore}, for the case $f_{4-10}=0.89$. This $f$
value represents a strongly attractive wall potential, as evidenced by
the high wall density. From the figure, one observes that at the
critical point the local density $\rho(z)$ is large compared to
$\rho_c$ across the whole width of the slit. As the temperature is
raised, however, the density in the slit middle decreases until, for
$T\gtrsim 1.09T_c$, it reaches the bulk value $\rho_c$.

The explanation of this behaviour is straightforward.  In the vicinity
of the critical point, the correlation length exceeds the slit width
$L$ and the density enhancement at the walls propagates across the
whole slit, raising $\rho(z)$  with respect to $\rho_c$. In this regime
one expects that for a sufficiently large slit width $L$, the density
would decay to its critical value like $z^{-\beta/\nu}$, as is the case
for critical adsorption at a single wall. Unfortunately our slit-pore
is much too narrow for $\rho(z)$ to reach the bulk value in the
available range of $z$. Neither do we observe pure power law scaling
for the variation that is present. This is because there exists no
substantial region of $z$ (away from the packing effects near the
walls) which is not simultaneously influenced by the potentials of both
walls.  

As the temperature is increased, there is a concomitant decrease in the
bulk correlation length $\xi\sim \tau^{-\nu}$ which at some point
becomes less than the slit width $L$.  For distances from the wall that
exceed $\xi$, the decay of $\rho(z)-\rho_c$ is expected to crossover to
an exponential form, $\exp(-z/\xi)$. We indeed observe a rapid relaxation
of $\rho(z)$ to $\rho_c$  for high temperatures ($T\gtrsim 1.09T_c$),
although for reasons similar to those described above, we have not been
able to identify its character.

\begin{figure}[h]  
\setlength{\epsfxsize}{8.0cm}
\centerline{\mbox{\epsffile{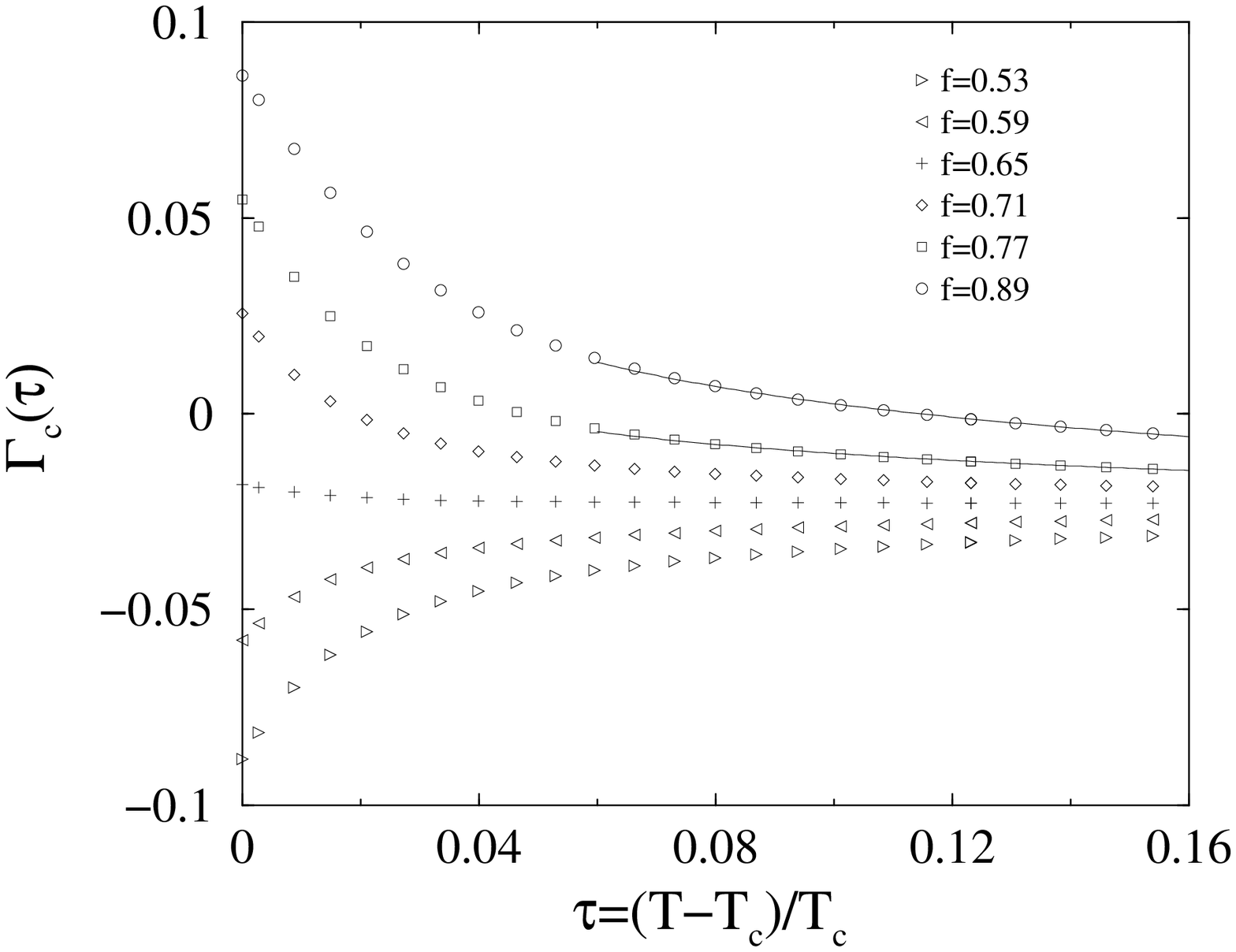}}}

\caption{The measured adsorption $\Gamma _c=\int_0^L (\rho(z)-\rho_c)dz$
( in units of $\sigma ^2$) for the $4$-$10$ potential on the critical isochore, plotted as a
function of the reduced temperature $\tau=(T-T_c)/T_c$. The slit width is
$L=20\sigma$.
Data is shown
for a selection of $f$ values, deriving from multi-histogram
reweighting \protect\cite{FERRENBERG88} of simulation data collected at
six temperatures in the range $T_c\leq T\leq 1.123T_c$. For the two
largest $f$ values we also show fits to the data in the large $\tau$
regime, of the form $\Gamma(\tau)=a+b\tau^{\beta-\nu}$ with
$\beta-\nu=-0.305$. Statistical errors do not exceed the symbol sizes.}

\label{fig:crit_ad}
\end{figure}

In order to quantify the temperature dependence of $\rho(z)$, and to
make contact with the experimental studies of
refs~\cite{THOMMES94,THOMMES95} and the theoretical results of
section~\ref{sec:denfun}, we have obtained the temperature dependence of
the adsorption, $\Gamma _c$, defined by Eq.(\ref{eq:gamc})
The form of $\Gamma _c(\tau)$ is shown in fig.~\ref{fig:crit_ad} for a
representative selection of $f$ values. The observed behaviour falls
naturally into $3$ regimes of $f$ values, namely large, intermediate and
small. 

For large $f$ ($f_{4-10}\geq 0.77$), fig.~\ref{fig:crit_ad} shows that
the adsorption increases monotonically as $\tau$ is reduced to zero from
above. Values of $f$ in this range are believed to correspond to the
situation studied experimentally in refs~\cite{THOMMES94,THOMMES95}.
As was the case  for DFT results of Sec.~\ref{sec:denfun}, we find no evidence for the experimentally observed depletion
phenomenon in which $\Gamma _c(\tau)$ first rises to a peak as $\tau$
decreases, and thereafter falls rapidly to negative values as the
critical point is approached.

We have attempted to analyse the form of $\Gamma _c(\tau)$ for large
$f$ at temperature well above criticality. Assuming
there exists a regime for which $L\gg\xi\gg\sigma$, (with $\sigma$ the
particle diameter), one expects that $\Gamma _c(\tau)$ will exhibit the universal
scaling behaviour of critical adsorption at a single wall, i.e.
$\Gamma _c(\tau) \sim \tau^{\beta-\nu}$. It is not clear, {\it a-priori},
that our rather narrow slit pore provides access to this regime.
Nevertheless fig.~\ref{fig:crit_ad} demonstrates that a fairly good
fit to this form can be achieved in the high temperature regime.

\begin{figure}[h]  
\setlength{\epsfxsize}{8.0cm}
\centerline{\mbox{\epsffile{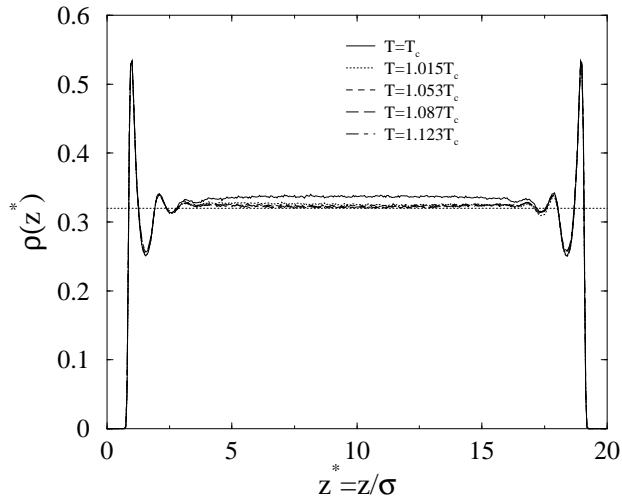}}}

\caption{The measured density profiles $\rho(z)$ ( in units of $\sigma ^3$)
for the $4$-$10$
potential with $f_{4-10}=0.65$ at a selection of temperatures along the
critical isochore of figure~\protect\ref{fig:isochore}. The
bulk density $\rho_c$ is denoted by a horizontal dotted line.}

\label{fig:neutral}
\end{figure}

As $f$ is reduced towards the intermediate regime, $\Gamma _c(\tau)$
becomes progressively flatter until, for $f_{4-10}\approx
0.65$, the adsorption appears to exhibit little or no temperature
dependence. This `neutral wall' scenario presumably arises from a
(near) cancellation of two competing factors associated with the wall
potential. On the one hand there is the `missing neighbours' effect
(represented by the parameter $c$ in equation~(\ref{eq:wallfluid})), whereby
particles close to the wall have their potential energy raised relative
to those in the bulk. On the other hand, there is the reduction in
potential energy of particles arising from the attractive part of
the wall potential (cf. the parameter $\epsilon _w$ in
equation~(\ref{eq:wallfluid})). The cancellation of these two contributions
effectively neutralises the influence of the wall on the fluid, and
with it the temperature dependence of $\Gamma _c$. This is also clearly
visible in the corresponding density profiles (fig.~\ref{fig:neutral}).

Turning finally to small $f$ values, we find that the adsorption is
negative at the critical point, reflecting the fact (cf.
fig.~\ref{fig:critprofs}) that $\rho(z)<\rho_c$ for all $z$. As $\tau$
increases, however, $\Gamma _c(\tau)$ is observed to increase
monotonically with increasing $\tau$, although it at no point becomes positive. This increase
in $\Gamma _c$ is traceable to a progressive relaxation (away from the
walls) of $\rho(z)$ towards the bulk value $\rho_c$. As in the case of
large $f$ discussed above, this effect has its origin in the decrease
of the correlation length.

\subsection{A sub-critical isochore}
In the experiments of refs~\cite{THOMMES94,THOMMES95}, the adsorption 
was studied on  sub-critical isochores having $0.995
\leq\rho/\rho_c\leq 0.999$. Unfortunately, in a simulation, it is not
practicable to employ these same isochores because of the wholesale
smearing-out of the critical region by finite-size effects.
Specifically, in the FSS regime, one finds \cite{WILDING92} that
$L^{\beta/\nu}(\rho-\rho_c)\simeq {\cal F} (u_HL^{d-\beta/\nu})$, where
${\cal F}$ is some scaling function. In practical terms, this means that
applying a given non-zero bulk field to the fluid engenders a much
smaller density change for a finite-size system than in the bulk limit.
As a corollary, one finds that the typical scale of density
fluctuations in a critical finite-sized system can be large. Thus for
instance, for a cubic simulation cell of size $L=17.5\sigma$, the
critical point density fluctuations extend from $\rho \approx 0.15$ to
$\rho\approx 0.5$. Clearly therefore a fractional reduction in the
density of $0.5\%$ would have little discernible effect, either on bulk
finite-size or adsorption phenomena.

In view of this, we have chosen to study $\Gamma _c(\tau)$ on the
isochore having $\rho=0.22\sigma ^{-3}=0.68\rho_c$. Although this density is considerably
smaller than $\rho_c$, it nevertheless lies within the range of
critical point density fluctuations in our finite-sized systems. One
can therefore anticipate that if there is a critical depletion effect
associated with negative values of the bulk field, then it should be
visible on this isochore.

The locus of the $\rho\sigma ^3=0.22$ isochore was determined in the temperature
range $T_c<T<1.123T_c$ using the same procedures as outlined in
section~\ref{sec:crit_iso}. However, an additional complication in the
present case was a greatly increased finite-size dependence of
$\mu_L(\rho,T)$. Comparison of $\mu_L(\rho,T)$ for $L=10\sigma$ and
$L=17.5\sigma$  revealed finite-size effects an order of magnitude
larger than those on the critical isochore (cf. inset of
fig.~\ref{fig:isochore}). This feature was discussed in
section~\ref{sec:crit_iso} and is traceable  to the differing
magnitudes of the bulk field $u_H$ on each isochore. To ameliorate the
problem, we determined $\mu(\rho,T)$ using a very large cubic system of
linear size $L=40\sigma$. Comparison of the results from $L=40\sigma$
with those from a system of size $L=17.5\sigma$ showed a discrepancy of
less than two parts in $10^4$.

\begin{figure}[h] 
\setlength{\epsfxsize}{8.0cm}
\centerline{\mbox{\epsffile{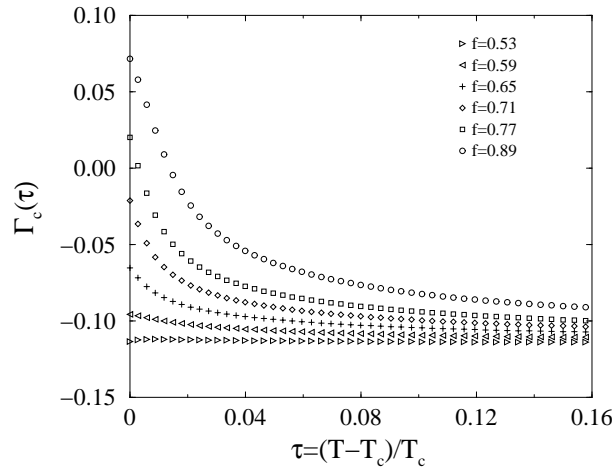}}}

\caption{The measured $\tau$ dependence of the adsorption
$\Gamma _c=\int_0^L (\rho(z)-\rho_c)dz$ ( in units of $\sigma ^2$)
for the $4$-$10$ wall potential on
the $\rho\sigma ^3=0.22$ isochore. Statistical errors do not exceed the symbol
sizes.}

\label{fig:gamma_sc}
\end{figure}

The form of $\Gamma _c(\tau)$ for the $4$-$10$ potential on the
$\rho\sigma ^3=0.22$
isochore is presented in fig.~\ref{fig:gamma_sc}. Data is shown for the
same selection of $f$ values given in fig.~\ref{fig:crit_ad}. Comparing
these two figures, it is clear that the application of the negative bulk
field reduces the adsorption to a considerably greater degree at large $\tau$
than at small $\tau$. We attribute this to the large
correlation length for small $\tau$, allowing the effect of the wall
density (which is effectively pinned by the choice of $f$) to propagate
across the slit, despite the action of the negative bulk field. The net
effect is to increase the range of variation of $\Gamma _c(\tau)$ for large
$f$ values but to reduce it for small $f$ values. The same qualitative
behaviour is observed for the $3$-$9$ wall potential.  However, in no
instance do we see any sign of a rapid critical depletion of the pore
density as $\tau\to 0$. 

\section{Discussion}
\label{sec:concs}

The density functional (DFT) results and those of simulation have 
confirmed the heuristic 
arguments of Section II that there should be no critical depletion 
for a fluid in a single
slit pore. In particular, for a fluid on its critical isochore, 
$\rho =\rho _c$, and for strongly
attractive walls the adsorption $\Gamma _c$ increases monotonically
 as $T$ is reduced towards $T_c$.
There is no evidence  of the rapid decrease of $\Gamma _c$ near 
$T_c$ which is the
experimentally observed signature of critical depletion. 
Although the simulation
results, Fig.8, do not display the saturation of $\Gamma _c$
 that is clear in DFT results of Fig.2
we believe  this reflects the different values of $L$ and the different
 temperature ranges
used in the two treatments. In the DFT calculations, where $L$ is very large, 
saturation does not set in until $\tau \sim 10^{-5}$ and $\Gamma _c$ 
is still decreasing
quite rapidly with $\tau$ for $\tau \ge 10^{-3}$. In the simulations, 
however, where 
$L=20\sigma $ we would not expect saturation until 
$\tau \sim \tau _0 \sim 3\times 10^{-3}$
and this is difficult to discern on the scale of Fig.8.
Such monotonic behaviour of $\Gamma _c$ is consistent
 with the lattice gas results for bulk field $H=0$~\cite{MACIOLEK98}.
It differs from the Monte Carlo results  of Schoen 
et.al.~\cite{SCHOEN95,SCHOEN97} but,
as mentioned earlier, those simulations suffered from systematic
 errors~\cite{WILDING99,NOTE2}.
The physical picture  which emerges from both our present 
treatments is one in  which the density
profiles show increasing enhancement, with respect to $\rho _c$, 
in the central portion of the slit
as the correlation length $\xi $ increases and the decay at each wall
 becomes slower.
When $\tau \sim \tau _0$, i.e. $\xi \sim L$, there is maximum enhancement 
and maximum
adsorption. Increasing the strength  of the wall-fluid  potential 
(the parameter $f$) 
simply raises the overall level of the profiles and leads to an increase in
$\Gamma _c$ at a given value of $\tau $ - see Fig.8.

For weakly attractive walls (small values of $f$) the simulation results yield
 negative values of the adsorption on the critical isochore, 
 with $\Gamma _c(\tau)$ 
monotonically increasing with increasing $\tau$. In this regime of $f$
the wall-liquid interface would be dry for $T<T_c$, i.e. the walls prefer gas 
to liquid. At supercritical temperatures the density near 
the wall is lower  than $\rho _c$ and the profile increases towards the centre 
of the slit. The closer one is to $T_c$ 
the slower is the increase and the more negative is $\Gamma _c$.
 Similar behaviour of
$\Gamma _c$ is found for very weakly attractive walls (small $\epsilon _w$) 
in our DFT
calculations and was noticed earlier by Marini Bettolo Marconi~\cite{MARINI88}.
By selecting an intermediate value of $f$ it is possible to construct a nearly 
'neutral wall' for which $\Gamma (\tau)$ is small in magnitude and almost
 independent    
of $\tau$. In this case the density profiles  are remarkably insensitive 
to $\tau $
 until very close  to $T_c$ where $\rho (z)$ has the striking form shown
  in Fig.9,
 i.e. it becomes concave upwards with respect to $\rho _c$.
 More careful analysis
  reveals that for  slightly greater values of $f$ the
 profiles  exhibit two symmetric maxima  near the walls
 (but outside the regions of the oscillatory structure)
 for temperatures close to $T_c$.
Similar, non-monotonic  profiles were found recently
for two-dimensional critical Ising
films in the crossover regime between ordinary and normal
transitions, i.e. for weak surface fields~\cite{MACIOLEK99}.
Currently we are investigating whether the non-monotonic  form of the profiles 
found in the present simulations of a Lennard-Jones fluid
arises from the same physical mechanism as in the Ising case.
 
We were motivated to investigate sub-critical isochores because 
i) the sorption experiments on porous glasses were performed 
 for $r\equiv (\rho-\rho _c)/\rho _c=-0.001$ and $-0.005$ and ii) the lattice
gas studies of Ref.\cite{MACIOLEK98,DRZEW98} 
 suggested that depletion should occur on a path at fixed bulk  field 
$\Delta \mu =2H<0$. Our present results for fluids show that no depletion
 occurs for
the wide range of sub-critical isochores and pore-widths which we have 
investigated.
The DFT results of Fig.2 imply that for the values of $r$ pertaining 
to the experiments
the behaviour of $\Gamma _c(\tau)$ is changed very little from that 
on the critical isochore $r=0$.
Increasing $\mid r\mid$ does alter $\Gamma _c(\tau)$ but this quantity
 remains monotonically increasing as $T\to T_c$. This is in keeping with
  the simulation
results (with large $f$) of Fig. 10, although it should be noted  that 
the simulations
refer to $r=-0.31$ and to a much smaller value of $L$ than in the DFT 
calculations
or in the experiments. In summary, our explicit calculations appear 
to confirm the expectations of Section II that, for strongly attractive walls
and parameter values of practical interest, the effective bulk field 
in the fluid
is insufficient to drive $\Gamma _c$ negative before $\xi \sim L$,
 i.e. we are in the regime $\tau _r\ll \tau _0$
where depletion does not occur. Alternatively one can say that for the relevant values of $r$,
 the experimental pore widths $L$ are not large enough for the density profile to relax from
its high value near the walls to its bulk sub-critical value in the centre.
Note that one cannot make $|r|$ too large or one leaves the experimental critical region.
The circumstances of the fluid are different from those of the lattice gas, 
where for 'reasonable' choices of fixed bulk field $H$ one
has~\cite{MACIOLEK98} $\tau _0 \ll \tau _H$, the parameter equivalent to $\tau
_r$.

We conclude that the adsorption calculated along critical and sub-critical
isochores for a single slit pore does not exhibit the depletion observed 
in the experiments. What then can be the explanation of the experimental data?
One might speculate that for some reason the density of the reference cell,
which fixes the experimental isochore, was not constant but it is
still difficult to see why  this might mimic the fixed $H$ scenario. What is
more likely is that the observed depletion arises from the fact that the porous
adsorbent is a complex solid material consisting of interconnected pores of
various shapes and sizes whose morphology is poorly understood.
Modelling such a material in terms of  ideal (non-connected) slit
pores is, of course, a gross over simplification. Recent theoretical and
simulation studies~\cite{REVIEWS}
show that the phase behaviour of fluids confined in such disordered media
{\it can be} very different from that which occurs  in a single pore
but we are not aware of investigations of the adsorption along near-critical
isochores. It should be possible to employ the model porous glasses 
produced in the quench simulations of Gelb and Gubbins~\cite{GELB},
which mimic the spinodal decomposition process used to make real CPG glasses,
in such an investigation. Were the adsorption  to be vastly different from that
found for the 'average' single  pore this would be a striking demonstration
of the importance of pore disorder and connectivity for critical phenomena.
Note that further evidence  that the single pore model  is inappropriate to
describe the experiments comes from the observation that critical depletion
occurs for the colloidal graphite substrate  Vulcan 3${\cal G}$ for $\rho/\rho _c=1.01$
and $1.04$ i.e. for super-critical isochores~\cite{THOMMES94,THOMMES95}.
Attempts to explain this observation within the single pore model require many further
assumptions\cite{MACIOLEK98}. 

\acknowledgements
A.M.  acknowledges beneficial discussions with Alina Ciach. She also thanks the Royal
Society/NATO  for financial support.
NBW acknowledges the financial support of the Royal Society (grant no.
19076), the EPSRC (grant no. GR/L91412) and the Royal Society of
Edinburgh.

$^1$ permanent address, Institute of Physical Chemistry, 
             Polish Academy of Sciences, Kasprzaka 44/52,
             PL-01-224 Warsaw, Poland

\appendix 
\section*{A}    
 For the Landau model  the integrals (\ref{eq:L})  and 
(\ref{eq:exgam}) can be performed explicitly  in the 
special case of $\Delta\mu=0$ (on the critical isochore $\rho _b=\rho
_c$)(see Ref.~\cite{MARINI88}):
\begin{equation}
\label{eq:Hsup}
L=2\rho_c^{-1/3}\left(\frac{D^*}{b^*}\right)^{1/2}(r_m^2+a^*/b^*)^{-1/2}F(\phi|m),
\end{equation}
where   $r_m=(\rho(L/2)-\rho_c)/\rho_c$ is the reduced density  at the midpoint
of the slit,
and $D^*\equiv  ( D\rho_c^{5/3})/k_BT_c$. 
$F(\phi|m)$ is the incomplete Jacobi elliptic
integral of the first kind~\cite{ABRAMOWITZ}  with arguments 
$\cos\phi=r_m/r_w$, where $r_w=(\rho(0)-\rho_c)/\rho_c$ is the reduced
density at the wall, and
$m=(r_m^2+2a^*/b^*)/(2r_m^2+2a^*/b^*)$.
On the
critical isochore $\rho_b=\rho_c$  the adsorption can also be evaluated 
~\cite{MARINI88}
\begin{equation}
\label{eq:gamsup}
\Gamma
_G=\rho_c^{2/3}\left(\frac{2D^*}{b^*}\right)^{1/2}\ln\left|\frac{r_w^2+\frac{a^*}{b^*}+\left[r_w^4-r_m^4+\frac{2a^*}{b^*}(r_w^2-r_m^2)\right]^{1/2}}{r_m^2+\frac{a^*}{b^*}}\right|
-\rho _cr_bL.
\end{equation} 
For $\Delta \mu <0$, i.e. for $\rho_b<\rho_c$, 
the relation (\ref{eq:L}) between  the order parameter at the midpoint $r_m$
and the wall separation $L$  
 can also be expressed  in terms of the elliptic
integral of the first kind.
For $T>T_c $, $a>0$ one can also show 
\begin{equation}
\label{eq:Lsub}
L=2\rho_c^{-1/3}\left(\frac{2D^*}{b^*}\right)^{1/2}(pq)^{-1/2}F(\phi|m),
\end{equation}
where  the arguments of   $F(\phi|m)$
are given by $\tan \phi/2=\left(\frac{q(r_w-r_m)}{p(r_w-r_1)}\right)^{1/2}$
and  $ m=(1/4)\frac{(p+q)^2+(r_m-r_1)^2}{pq}$.
$r_1$ is the single real root of  the cubic equation
\begin{equation}
\label{eq:cubeq}
r^3+r^2r_m+r(r_m^2+2a^*/b^*)+r_m^3+(2a^*/b^*)r_m-(4/b^*)\Delta\mu^*=0
\end{equation}
 and
$p^2=(u-r_m)^2+n^2$, $q^2=(u-r_1)^2+n^2$ where 
$u=-(1/2)(r_1+r_m)$ and $n^2=r_m^2-u^2-2ur_1+2a^*/b^*$.

Also for the case of Fisk-Widom free energy, the integrals in the formulae for $L$ and $\Gamma _G$
 can be  expressed in terms of elliptic integrals of the first kind
 for the special case of the critical isochore $\Delta \mu=0$:
 \begin{equation}
 \label{eq:LWid}
 L=\left(\frac{3D^*}{b^*y}\right)^{1/2}\frac{1}{\lambda}F(\phi\setminus(\pi/2-\alpha))
 \end{equation}
 where $y=r_m^2((3a^*/b^*)+r_m^4)$ and
 \begin{equation}
 \label{eq:lam}
 \lambda^2=(\frac{-6(a^*/b^*)}{yr_m^2}+3r_m^{-4})^{1/2},
 \end{equation}
 \begin{equation}
 \label{eq:cs1}
 \cos\phi=(\lambda^2-\frac{1}{r_m^2}+\frac{1}{r_w^2})/(\lambda^2+\frac{1}{r_m^2}-\frac{1}{r_w^2})
\end{equation}
\begin{equation}
\label{eq:sin1}
\sin^2\alpha=1/2-(3/4)\frac{r_m^{-2}-a^*/(b^*y)}{\lambda^2}
\end{equation}
and
\begin{equation}
 \label{eq:gWid}
 \Gamma
 _G=\left(\frac{3D^*}{b^*}\right)^{1/2}\rho_c\frac{1}{\lambda}F(\phi\setminus\alpha)-\rho
 _cr_bL
 \end{equation}
 where 
 \begin{equation}
 \label{eq:lam2}
 \lambda^2=(3(a^*/b^*)+3r_m^4)^{1/2},
 \end{equation}
 \begin{equation}
 \label{eq:cs2}
 \cos\phi=(\lambda^2-r_w^2+r_m^2)/(\lambda^2+r_m^2-r_w^2)
\end{equation}
\begin{equation}
\label{eq:sin2}
\sin^2\alpha=1/2-\frac{3}{4}(r_m^2/\lambda^2)
\end{equation}
Equivalent forms were derived in  Ref.~\cite{MARINI88} where they  were
  used to study the 
 behaviour of the adsorption in the crossover
 regime  from the noncritical  to the scaling region  as a function of wall
 separation and  surface fields.


\begin{thebibliography}{99}
\bibitem{NOTE} Note that this definition of the dimensionless $\Delta \mu^*$
differs slightly from that used in Sec.II for the vdW fluid.
\end{thebibliography}
\end{document}